\documentclass[preprint,12pt,a4paper]{elsarticle}

\usepackage{multirow}

\usepackage{epsfig}

\usepackage{amssymb}

\biboptions{compress}

\begin{document}

\begin{frontmatter}

\title{Exploring central opacity and asymptotic scenarios in elastic hadron scattering}

\author[label1]{D. A. Fagundes}
\author[label2]{M. J. Menon}
\author[label2]{P. V. R. G. Silva}

\address[label1]{Universidade Federal de Santa Catarina - Campus Blumenau,
Rua Pomerode 710, Salto do Norte 89065-300
Blumenau, SC, Brazil}
\address[label2]{Instituto de F\'{\i}sica Gleb Wataghin,
Universidade Estadual de Campinas - UNICAMP\\
13083-859 Campinas, SP, Brazil}

\begin{abstract} 
In the absence of a global description of the experimental data on elastic and
soft diffractive scattering from the first principles of QCD, model-independent
analyses may provide useful phenomenological insights for the 
development of the theory in the soft sector. With that in mind, 
we present
an empirical study on the energy dependence of the ratio
$X$ between the elastic and total cross sections;
a quantity related to the evolution of the hadronic central opacity. 
The dataset comprises all the experimental information available on proton-proton 
and antiproton-proton scattering in the c.m energy interval 5 GeV - 8 TeV. 
Generalizing previous works, we discuss four model-independent analytical parameterizations 
for $X$, consisting of  
sigmoid functions composed with elementary functions of the energy
and three distinct asymptotic
scenarios: either the standard black disk limit or scenarios above or 
below that limit. Our two main conclusions are the following: (1) although consistent with the 
experimental data, the black disk does not represent an unique solution;
(2) the data reductions favor a semi-transparent scenario, with asymptotic
average value for the ratio $\bar{X}$ = 0.30 $\pm$ 0.12.  
In this case, within the uncertainty, the asymptotic regime may already be reached around 1000 TeV.
We present a comparative study of the two scenarios, including predictions 
for the inelastic channel (diffraction dissociation)
and the ratio associated with the total cross-section
and the elastic slope.
Details on the selection of our  empirical ansatz for $X$ and  physical
aspects related to a change of curvature in this quantity at 80 - 100 GeV,
indicating the beginning of a saturation effect, are also 
presented and discussed.
\end{abstract}
 
\begin{keyword}
Hadron-induced high- and super-high-energy interactions
\sep
Total cross-sections
\sep
Asymptotic problems and properties
13.85.-t
\sep
13.85.Lg
\sep
11.10.Jj

\end{keyword}

\end{frontmatter}

\vspace{1.5cm}

\centerline{\textit{Published in Nuclear Physics A 946 (2016) 194-226}}

\newpage

\centerline{\textbf{Table of Contents}}


1. Introduction


2. Experimental Data and Asymptotic Scenarios

\ \ \ \ 2.1 Dataset

\ \ \ \ 2.2 Asymptotic Scenarios

\ \ \ \ \ \ 2.2.1 The Black Disk

\ \ \ \ \ \ 2.2.2 Above the Black Disk

\ \ \ \ \ \ 2.2.3 Below the Black Disk


3. Model-Independent Parametrization

\ \ \ \ 3.1 Empirical and Analytical Arguments

\ \ \ \ \ \  3.1.1 General Aspects

\ \ \ \ \ \  3.1.2 Sigmoid Functions $S(f)$

\ \ \ \ \ \  3.1.2 Elementary Functions $f(s)$

\ \ \ \ 3.2 Analytical Parameterizations and Notation

\ \ \ \ 3.3 The Energy Scale

\ \ \ \ 3.4 Constrained and Unconstrained Fits


4. Fit Procedures and Results 

\ \ \ \ 4.1 Fit Procedures

\ \ \ \ 4.2 Fit Results with the Logistic

\ \ \ \ \ \ 4.2.1 Variant $PL$

\ \ \ \ \ \ 4.2.2 Variant $LL$

\ \ \ \ 4.3 Fit Results with the Hyperbolic Tangent

\ \ \ \ \ \ 4.3.1 Variant $PL$

\ \ \ \ \ \ 4.3.2 Variant $LL$

5. Discussion and Conclusions on the Fit Results

\ \ \ \ 5.1 Constrained and Unconstrained Fits

\ \ \ \ 5.2 Logistic and Hyperbolic Tangent

\ \ \ \ 5.3 Variants $PL$ and $LL$

\ \ \ \ 5.4 Inflection Point

\ \ \ \ 5.5 Conclusions on the Fit Results

\ \ \ \ \ \ 5.5.1 Main Conclusions

\ \ \ \ \ \ 5.5.2 Selected Results and Scenarios


6. Extension to Other Quantities 

\ \ \ \ 6.1 Inelastic Channel: Ratios and Diffractive Dissociation

\ \ \ \ 6.2 Ratio Y Associated with Total Cross-Section and Elastic Slope


7. Further Comments

\ \ \ \ 7.1 On Possible Physical Interpretations

\ \ \ \ 7.2 On a Semi-Transparent Scenario


8. Summary, Conclusions and Final Remarks


Appendix A. Basic Formulas and Results

\ \ \ \ A.1 Impact Parameter and Eikonal Representations

\ \ \ \ A.2 The $X$ and $Y$ Ratios

Appendix B. Short Review of Previous Results

\newpage

\section{Introduction}
\label{s1}

Run 1 of the CERN-LHC on $pp$ collisions has revealed 
important new features of the strong interactions at the highest energy region
reached by accelerator machines (7 TeV - 8 TeV in the c.m. system) and
excitement grows with the start of Run 2 at 13 TeV.
However, despite all the theoretical successes of the Standard Model,
the \textit{soft strong interactions} (small momentum transfer) still constitute a great 
challenge for QCD. Once related to large distance phenomena, they are not formally accessed 
by perturbative
techniques and the crucial point concerns
the absence of a nonperturbative framework able to provide \textit{from the first principles} of QCD
a global description of
the soft scattering states,
namely the experimental data on elastic and inelastic diffractive processes (single and double dissociation)
and that constitutes a long-standing problem
\cite{pred,land}.

A renewed interest in these processes has been
due to the large amount of experimental data
on \textit{elastic scattering} (including total cross-section) that has been provided  
by the TOTEM Collaboration at 7 and 8 TeV
\cite{totem1,totem2,totem3,totem4}, more recently by the ATLAS Collaboration 
at 7 TeV \cite{atlas} and also due to the present expectations with Run 2 at 13 TeV \cite{forward}. 
Beyond the accelerator energy region, cosmic-ray experiments constitute the only tool for
investigating particle properties and their interactions. Studies
on extensive air showers (EAS) \cite{engel1} allow the determination of the 
proton-air production cross-section and from this quantity, it is
possible to estimate the proton-proton cross-section \cite{engel2,engel3} at energies above 50 TeV.
However, in practice, the interpretation of the EAS development depends on
extrapolations from theoretical formalisms that have been tested only in the
accelerator energy region, resulting in rather large theoretical uncertainties,
mainly in the estimation of the total cross-section.

In the absence of pure QCD descriptions, the theoretical investigation of the soft strong scattering is 
still restricted to a phenomenological context and therefore characterized by some
intrinsic limitations. 
In fact, although partial descriptions of the bulk of the experimental data can
be obtained in the context of phenomenological models (for reviews see, for example, 
\cite{dremin,godizov,cartiglia,kaspar,fiore,matthiae}),  
the efficiency of any representative approach depends on a constant feedback of new and adjustable 
parameters, which
in turn are dictated by the new 
experimental data.
As a practical consequence, equivalent data descriptions may be obtained
in different phenomenological contexts, associated, in general, to \textit{different physical 
pictures for the same 
phenomenon} (compare, for example, 
\cite{kmr,glm,glm-sat,bdh,nemes,selyugin,dl13,kpp,grau,wibig,dgm,dgm05}), resulting, nearly always,
in a renewed open problem.

At this stage, another important strategy is the development of  \textit{model independent} 
descriptions of the
experimental data, looking for quantitative \textit{empirical} results 
that may work as an effective bridge for further developments of the QCD in the soft 
scattering sector and even selecting phenomenological pictures.
As discussed in what follows, that is the point we are interested in here.

Among the physical quantities characterizing the elastic hadron-hadron scattering,
the ratio $X$ between the elastic (integrated) and total cross sections
as a function of the c.m. energy $\sqrt{s}$,
\begin{eqnarray}
X(s) = \frac{\sigma_{el}}{\sigma_{tot}}(s),
\label{x}
\end{eqnarray}
plays an important role for several reasons.

\begin{description}

\item{1.} 
In the impact parameter representation ($b$-space), it is connected with the
opacity (or blackness) of the colliding hadrons; as we shall recall,
it is proportional to the central opacity ($b$ = 0) in the cases of the gray/black disk
or Gaussian profiles. 

\item{2.} 
Due to model-dependence involved in direct measurements of the
inelastic cross section, the $s$-channel unitarity  constitutes the less biased
way to obtain $\sigma_{inel}$, namely from $\sigma_{tot} - \sigma_{el}$. Therefore,  
the same is true for the ratio $\sigma_{inel}/\sigma_{tot} = 1 - X$ which, in turn, 
can be associated with the inelasticity of the 
collision \cite{jdd}.

\item{3.} 
Empirical information on $X(s)$ 
and $1 - X(s)$ at the highest $s$ and as $s \rightarrow \infty$,
provides crucial information on the asymptotic properties of the hadronic interactions,
namely information on how $\sigma_{tot}$ and  $\sigma_{el}$ reach simultaneously their
respective unitarity bounds (which is one of the main prospects in the
LHC forward physics program \cite{forward}). 
Empirical information on asymptopia is also important in the construction and/or selection of 
phenomenological models, mainly those based or inspired in 
nonperturbative QCD. 

\item{4.} 
Through an approximated relation (to be discussed later), the ratio $X$ can be  connected
with the ratio between the total cross-section and the elastic slope parameter ($B$),
\begin{eqnarray}
Y(s) = \frac{1}{16 \pi}\frac{\sigma_{tot}}{B}(s).
\label{y}
\end{eqnarray}
This ratio, extrapolated to cosmic-ray energies, plays an important role
in the study of extensive air-showers, as will be discussed along the paper. Moreover,
it gives also information on the connection between $\sigma_{tot}$ and $B$
at the highest and asymptotic energies.
    
\end{description}

Based on these introductory remarks, our goal here is to present an empirical analysis
on the ratio $X(s)$ and discuss the implications of the results along the
aforementioned lines, with main focus on asymptotic scenarios ($s \rightarrow \infty$).
In this respect, previous results obtained with smaller sized datasets and treating particular aspects, have already
been reported in \cite{fm12,fm13,fms15a,fms15b}; we shall review these results later.

In the present work, our empirical study on the above quantities is once more updated, developed
and extended in several aspects. 
The dataset on $X$ comprises $pp$ and $\bar{p}p$ scattering above 5 GeV, including all 
the TOTEM data at 7 and 8 TeV and also the recent ATLAS datum at 7 TeV.
Generalizing previous works, we introduce four analytical parameterizations for $X(s)$,
consisting of two sigmoid functions composed with two elementary functions of the energy.
Arguments and procedures concerning the selection of these suitable empirical parameterizations
are presented and discussed in detail. 
As in \cite{fms15a,fms15b},
we investigate all the three possible asymptotic scenarios: either the standard black disk limit or
scenarios above or bellow that limit. 
Depending on the case and variant considered, the parametrization has only 
three to five free fit parameters. 
Based on the $s$-channel unitarity, the results for $X(s)$ are extended to the inelastic channel,
with discussions on the dissociative processes (single, double diffraction), including the Pumplin bound. 
We also treat the connection between
$X(s)$ and the ratio $Y(s)$, Eq. (\ref{y}), together with discussions
on the applicability of the results in studies of extensive air showers in cosmic-ray experiments.

On the basis of the dataset, empirical parameterizations, fit procedures and results, our main four conclusions 
are the following:
(1) although consistent with the experimental data, the black disk limit does not represent an unique 
or exclusive solution;
(2) with the asymptotic $X$-value as a free fit parameter, all the data reductions
favor a scenario below the black disk, with estimated average result 0.30 $\pm$ 0.12;
(3) within the uncertainties our selected parameterization and fit result indicate
that asymptopia may already be reached  around 10$^3$ TeV;
(4) a change from positive to negative curvature in $X(s)$ is
predicted at 80 - 100 GeV, which suggests a change in the dynamics
of the strong interactions associated with the beginning of a saturation effect.

The article is organized as follows.
Our analysis on the ratio $X$ is developed through Sections \ref{s2} to \ref{s4}, where
we present: the dataset investigated and the arguments why to treat 
also limits below and 
above the black disk (Section \ref{s2});
a  detailed discussion on the selection of our empirical 
parametrization and variants considered (Section \ref{s3}); 
the fit procedures and the fit results (Section \ref{s4}).
In Section \ref{s5} we discuss in a comparative way
all the data reductions and present our conclusions on the fit results,
which lead to the selection of two asymptotic scenarios: semi-transparent and black.
In Section \ref{s6}, we display the predictions for some  quantities associated with the
inelastic channel 
and  the ratio between the total cross-section and the elastic slope. 
In Section \ref{s7}, we discuss some possible physical aspects
related to the analytical structure of the empirical parameterizations,
presenting also comments on a semi-transparent asymptotic scenario.
A summary and our final conclusions are the contents of  
Section \ref{s8}. In two appendixes, we present some basic formulas and results
referred to along the text (\ref{saa}) and a short review of our previous works
on the subject  (\ref{sab}).

\section{Experimental Data and Asymptotic Scenarios}
\label{s2}

\subsection{Dataset}

Our dataset comprises all the experimental information presently available 
on the ratio $X$ from $pp$ and $\bar{p}p$
scattering above 5 GeV and up to 8 TeV (42 points: 29 from $pp$
and 13 from $\bar{p}p$) and have been collected from \cite{pdg}. The statistical and systematic
uncertainties have been added in quadrature. 

\begin{figure}[h!]
\centering
\epsfig{file=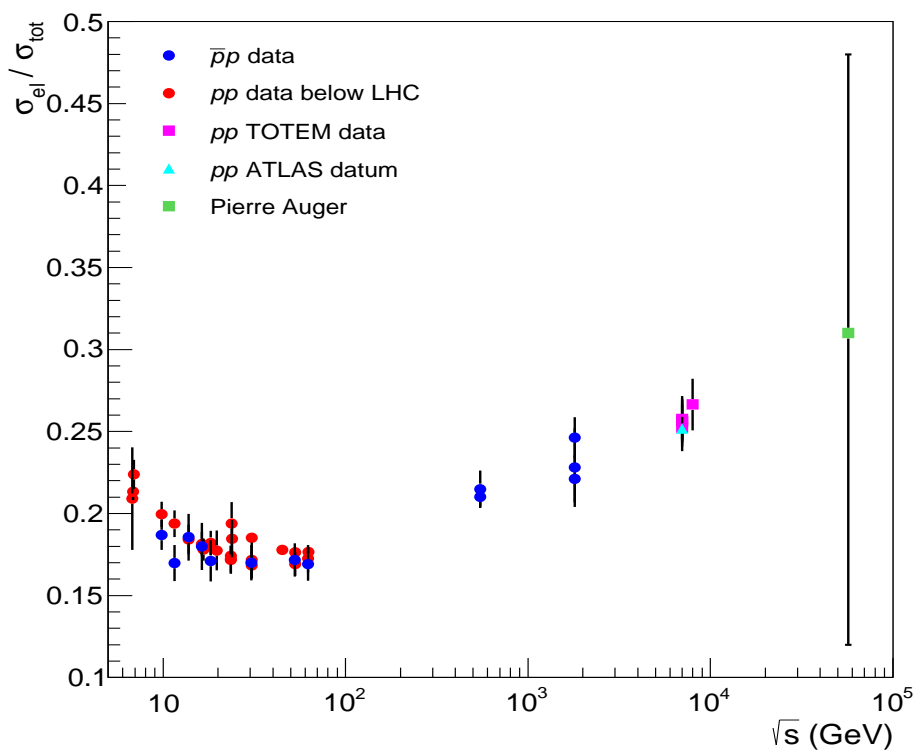,width=12cm,height=10cm}
\caption{Experimental data on the ratio $X=\sigma_{el}/\sigma_{tot}$ used in this analysis, from $pp$ and $\bar{p}p$
elastic scattering in the energy region 5 GeV $< \sqrt{s} \leq$ 8 TeV (accelerator data). 
At $\sqrt{s}$ = 57 TeV it is shown, as illustration, the ratio obtained from the estimations
of $\sigma_{tot}$ and $\sigma_{inel}$ in cosmic-ray experiment (Pierre Auger Observatory).}
\label{f1}
\end{figure}

As we will show, the $pp$ data at 
the LHC region play a crucial role in our analysis and conclusions. 
These 6 points are displayed in Table \ref{t1}:
four points at 7 TeV, one point at 8 TeV obtained by the TOTEM Collaboration and one point at 7 TeV
recently obtained by the ATLAS Collaboration.  
For further reference, a plot of the above-mentioned dataset is presented in Fig. \ref{f1}. 
As illustration, we have also included an estimation of the ratio $X$ at 57 TeV,
evaluated from the experimental information on $\sigma_{tot}$ and $\sigma_{inel}$
obtained by the Pierre Auger Collaboration \cite{auger}. This value, that did not
take part of our data reductions, reads 
\begin{eqnarray}
X(\sqrt{s} = 57\, \mathrm{TeV}) = 0.31 ^{+0.17}_{-0.19},
\nonumber
\end{eqnarray}
where statistical, systematic and Glauber uncertainties \cite{auger}
have been added in quadrature.

\begin{table}[ht]
\centering
\caption{Experimental data on $\sigma_{tot}$, $\sigma_{el}$ and the corresponding 
ratio $X=\sigma_{el}/\sigma_{tot}$ at the
LHC energy region.}   
\vspace{0.1cm}
\begin{tabular}{c c c c c}
\hline
$\sqrt{s}$ & $\sigma_{tot}$ & $\sigma_{el}$ & $\sigma_{el}/\sigma_{tot}$ & Collaboration  \\
 (TeV)    &     (mb)       &   (mb)        &                            &                 \\
\hline
7 & 98.3  $\pm$ 2.8 & 24.8 $\pm$ 1.2 & 0.252 $\pm$ 0.014 & TOTEM \cite{totem1} \\
7 & 98.6  $\pm$ 2.2 & 25.4 $\pm$ 1.1 & 0.258 $\pm$ 0.013 & TOTEM \cite{totem2} \\
7 & 98.0  $\pm$ 2.5 & 25.1 $\pm$ 1.1 & 0.256 $\pm$ 0.013 & TOTEM \cite{totem3} \\
7 & 99.1  $\pm$ 4.1 & 25.4 $\pm$ 1.1 & 0.256 $\pm$ 0.015 & TOTEM \cite{totem3} \\
7 & 95.4  $\pm$ 1.4 & 24.0 $\pm$ 0.6 & 0.252 $\pm$ 0.004 & ATLAS \cite{atlas}  \\
8 & 101.7 $\pm$ 2.9 & 27.1 $\pm$ 1.4 & 0.266 $\pm$ 0.016 & TOTEM \cite{totem4} \\
\hline 
\end{tabular}
\label{t1}
\end{table}

\subsection{Asymptotic Scenarios}

We present here some theoretical, phenomenological and empirical arguments that led
us to investigate all the three possible asymptotic scenarios. 
Let us denote $A$ the
asymptotic value of the ratio $X$:
\begin{eqnarray}
\lim_{s \rightarrow \infty} X(s) \equiv A.
\label{a}
\end{eqnarray}
In what follows we discuss and display some numerical values for $A$,
which will be considered as typical
of each scenario to be investigated in Section \ref{s4}, through data reductions.

\subsubsection{The Black Disk}

As recalled in \ref{saa}, the black disk limit reads
\begin{eqnarray}
A = \frac{1}{2}.
\nonumber
\end{eqnarray}
This limit represents a standard phenomenological
expectation. It is typical of eikonalized formalisms, as the traditional models by
Chou and Yang \cite{cy}, Bourrely, Soffer and Wu \cite{bsw1,bsw2,bsw3}, the hybrid approach by
Block and Halzen \cite{bhbd} and a number of models that have been
continuously refined and developed 
(for example, \cite{kmr,glm,glm-sat,bdh,nemes,selyugin,grau,wibig,dgm}).

\subsubsection{Above the Black Disk}

We have the arguments that follows for investigating scenarios above
(beyond) the black disk.

\begin{description}

\item{1.} In the formal context, the $s$-channel unitarity,
\begin{eqnarray}
\frac{\sigma_{el}}{\sigma_{tot}} + \frac{\sigma_{inel}}{\sigma_{tot}} = 1,
\label{unit}
\end{eqnarray}
imposes an obvious maximum bound for $X(s)$, namely
\begin{eqnarray}
A = 1.
\nonumber
\end{eqnarray}

\item{2.}
In the phenomenological context, the $U$-matrix unitarization scheme by Troshin and Tyurin
predicts an asymptotic limit beyond the black disk
\cite{tt1,tt2}; in this approach,
the reflective scattering mode becomes active at small values of the impact parameter \cite{amkt}.

\item{3.}
In the formal context, two well known bounds have been
established for the total cross-section \cite{froissart,lukmar} and the
inelastic cross-section \cite{martin,martinwu},
\begin{eqnarray}
\sigma_{tot}(s) < \frac{\pi}{m_{\pi}^2} \ln^2(s/s_0)
\quad \mathrm{and} \quad
\sigma_{inel}(s) < \frac{\pi}{4m_{\pi}^2} \ln^2(s/s_0).
\nonumber
\end{eqnarray}
Therefore, in case of simultaneous saturation of both bounds
as $s \rightarrow \infty$, it is possible that 
$\sigma_{inel}/\sigma_{tot} \rightarrow 1/4$, which from unitarity, Eq. (\ref{unit}), implies in
\begin{eqnarray}
A = \frac{3}{4}.
\nonumber
\end{eqnarray}
It should be noted that this limit does not correspond to an usual interpretation of
the aforementioned asymptotic bounds. In fact, two fractions $1/2$ are usually associated 
with each bound, in place of
$1/4$ in the inelastic case, which favors the black disk scenario \cite{dremin,cartiglia,martin,bh2011}. Even if a simultaneous
saturation of both bounds might be questionable \cite{ttbound}, the number 0.75 can be considered
as an instrumental choice for data reductions, lying between the black disk 
and the maximum value allowed by unitarity.
\end{description}

\subsubsection{Below the Black Disk}

Nonetheless seeming a rather unorthodox possibility, in what follows we recall some results 
which suggest and indicate asymptotic limits below the black disk.

\begin{description}

\item{1.}
In the publications of the TOTEM Collaboration, the authors quote the prediction
for the total cross-section obtained by the COMPETE Collaboration (in 2002), 
with parametrization \textit{RRPL2}, energy cutoff at 5 GeV and given by \cite{compete}
\begin{eqnarray}
\sigma_{tot}^{COMPETE}(s) = 42.6\, s^{-0.46} - 33.4\, s^{-0.545} + 35.5 + 0.307 \ln^{2}(s/29.1),
\label{comp}
\end{eqnarray}
where all coefficients are in mb and $s$ is in GeV$^2$.
In the same publications, a fit to the $\sigma_{el}$ data above 10 GeV by the TOTEM Collaboration
is also presented \cite{totem4},
\begin{eqnarray}
\sigma_{el}^{TOTEM}(s) = 11.7 - 1.59 \ln(s) + 0.134 \ln^{2}(s).
\label{tot}
\end{eqnarray}

Using the above parameterizations, the ratio $X$ for $s \rightarrow \infty$ reads
\begin{eqnarray}
A =  0.436,
\nonumber
\end{eqnarray}
which suggests a scenario below the black disk.
We shall return to parameterizations Eqs. (\ref{comp}) and (\ref{tot}) in the next sections.

\item{2.} 
We have recently developed several analyses of the experimental data on
$\sigma_{tot}$, the $\rho$ parameter and $\sigma_{el}$ \cite{fms13,ms13a,ms13b}, including the TOTEM
Collaboration results at 7 and 8 TeV. 
For our purposes, we recall that the parametrization
for the total cross section is expressed by
\begin{eqnarray}
\sigma_{tot}(s) = \mathrm{Regge}\ \mathrm{terms}\ + \alpha + \beta \ln^{\gamma}(s/s_h)
\nonumber
\end{eqnarray}
where $\alpha$, $\beta$ and $\gamma$ are free parameters.
Fits to data on $\sigma_{tot}$ and $\rho$ (using derivative dispersion relations) from
$pp$ and $\bar{p}p$ scattering above 5 GeV, have led to statistically consistent solutions
either with $\gamma = 2$ (fixed) or  $\gamma > 2$ (free fit parameter).
In \textit{both cases}, extension of the parametrization to $\sigma_{el}$ data (same $\gamma$ value) 
allowed to extract the ratio $X(s)$ and its asymptotic value $A$. In all cases we have obtained
$A < 1/2$, with lowest central value around $1/3$ (see a summary of the results in \cite{ms13b},
figure 10, where the above point obtained with the TOTEM and COMPETE results is also displayed).
For future use, as a typical input in data reductions, we shall consider the lowest value obtained
in these analyses, which, within the uncertainties, reads
\begin{eqnarray}
A = 0.3.
\nonumber
\end{eqnarray}

\end{description}

\section{Model-Independent Parametrization}
\label{s3}

As we shall demonstrate in the next sections, our model-independent parameterizations
for $X(s)$ constitute useful, practical and efficient tools in both the description
of the experimental data and the extensions to other physical quantities. Given this
important fact, we present in this section a detailed
discussion on the choices and steps that led us to the construction of the empirical
ansatz for $X(s)$. 

\subsection{Empirical and Analytical Arguments}

\subsubsection{General Aspects}

First, from Fig. \ref{f1}, as the energy increases above 5 GeV, the $X$ data decreases
up to the CERN-ISR region ($\approx$ 20 - 60 GeV), where they remain approximately constant and then
begin to increase smoothly. From a strictly empirical point of view, this rise in the linear-log
plot scale may suggest a parabolic parametrization in terms of $\ln s$, that is, an increasing
slope (positive curvature). However, as discussed in Section  2.2, in the formal context, 
the Unitarity Principle demands the obvious bound $1$ for $X(s)$ as
$s \rightarrow \infty$ and finite values
are also dictated by the Froissart Martin bounds and all phenomenological
models and empirical analysis, independently of the scenario (black disk,
above or below).
Therefore, except in case of existence of an unexpected singular behavior at
some finite value of the energy, the above facts indicate a constant finite
asymptotic limit for the ratio $X(s)$, that is, a smooth \textit{saturation effect}
as $s \rightarrow \infty$.
That in turn, demands a change in the sign of the curvature at some finite value
of the energy, so that $X(s)$ goes asymptotically to a constant limit with
a decreasing slope (negative curvature).

The above empirical and formal arguments concerning $X(s)$ 
suggest an analytical parameterization
related to a sigmoid function (``S-shaped" curves), in order to impose the change of 
the curvature sign and an asymptotic limit (a constant).  
However, there is one more step, because the description of
the data at low energies and the correct curvature dictated by the experimental data, 
depend also on a suitable choice for the functional argument of the sigmoid function.
That led us to first express the ratio as follows.
Denoting $S$ a sigmoid function and $f$ its argument depending on $s$, we  express
$X$ as a composite function,
\begin{eqnarray}
X(s) = A\,S(f), \qquad f = f(s),
\label{xafs}
\end{eqnarray}
where, from Eq. (\ref{a}),
\begin{eqnarray}
\lim_{s \rightarrow \infty} S(f(s))= 1.
\nonumber
\end{eqnarray}

Therefore and still on empirical grounds, the point would be to test different forms 
for $S(f)$ and $f(s)$ through
fits to the experimental data, looking for statistically consistent descriptions
with an \textit{economical number} of free parameters.
Obviously the problem does not have a unique solution and it is not possible
to test all the analytical possibilities. 

However, the use of different classes of sigmoid functions combined with different classes
of phenomenologically based elementary functions is a way to take into account the 
intrinsic uncertainty in the choice of the complete parametrization.
With that in mind,
we shall select two forms for $S(f)$, combined (each one) with two forms for $f(s)$,
as explained in what follows.

\subsubsection{Sigmoid Functions $S(f)$}

Several classes of sigmoid functions have applications in different scientific contexts
and that includes the logistic, hyperbolic tangent, error function, algebraic ratios 
and many others (see for example \cite{sig1,sig2}; we shall return to its applications
in Section 7.1). 
Here we consider two classes of sigmoid functions. One of them, already used in all our
previous analyses \cite{fm12,fm13,fms15a,fms15b}, is the \textit{Hyperbolic Tangent}, denoted by
\begin{eqnarray}
S^{HT}(f) = \tanh f = \frac{1 - \exp\{{-2f}\}}{1 + \exp\{{-2f}\}}.
\label{ht}
\end{eqnarray}

In addition,  we now consider a \textit{Logistic} function, with notation
\begin{eqnarray}
S^L(f)  = \frac{1}{1 + \exp\{-f\}},
\label{l}
\end{eqnarray}
which can be connected with $S^{HT}(f)$ by translation and scaling transformation. 

\subsubsection{Elementary Functions $f(s)$}

We shall express $f(s)$ in terms of elementary
functions of the \textit{standard soft variable} denoted
\begin{eqnarray}
\ln{(s/s_0)} \equiv v,
\label{v}
\end{eqnarray}
where $s_0$ is a fixed energy scale to be discussed later.
Different functions and different conditions have already been
investigated in our previous analyses \cite{fm12,fm13,fms15a,fms15b}.
Here, in order to generalize our previous parametrization and extend
our analysis, we first express $f$ as a sum of two terms: a linear function of the
standard variable $v$ and a function $g(v)$ which, dictated by the data reductions, can account for
possible deviations from linearity:
\begin{eqnarray}
f(s) \rightarrow f(v) = \alpha + \beta\, v + \gamma\, g(v),
\label{fs}
\end{eqnarray}
where $\alpha$, $\beta$ and $\gamma$ are real free fit parameters.

Different tests, using distinct datasets
(only $pp$ or including $\bar{p}p$) and different energy cutoffs, under the above
mentioned criteria, led us to choose two typical forms in soft scattering for $g(v)$,
namely either a \textit{Power-Law},
\begin{eqnarray}
g_{PL}(v) = v^{\delta},
\label{pl}
\end{eqnarray}
where $\delta$ is an (additional) free fit parameter, or a \textit{Logarithmic-Law},
\begin{eqnarray}
g_{LL}(v) = \ln v,
\label{ll}
\end{eqnarray}
therefore, without additional parameter.
 
Here ends the arguments and steps used in the analytical construction of our empirical 
parameterizations (or ansatz) for $X(s)$, which is summarized and denoted
in the next subsection.

\subsection{Analytical Parameterizations and Notation}

Collecting and summarizing  Eqs. (\ref{xafs}) to (\ref{ll}) and as a matter of  
notation, we shall express our four choices of parameterizations as follows.
For each sigmoid function (logistic or hyperbolic tangent) we consider
two \textit{variants} associated and denoted by the functions $g(v)$.

\begin{description}

\item[-]
Logistic ($L$) with variant Power-Law ($PL$)

\begin{eqnarray}
X_{PL}^{L}(s) =  \frac{A}{1 + \exp\{-[\alpha + \beta \ln (s/s_0) + \gamma \ln^{\delta} (s/s_0)]\}};
\label{lpl}
\end{eqnarray}

\item[-]
Logistic ($L$) with variant Logarithmic-Law ($LL$)

\begin{eqnarray}
X_{LL}^{L}(s) =  \frac{A}{1 + \exp\{-[\alpha + \beta \ln (s/s_0) + \gamma \ln \ln (s/s_0)]\}}.
\label{lll}
\end{eqnarray}

\item[-]
Hyperbolic Tangent ($HT$) with variant Power-Law ($PL$)

\begin{eqnarray}
X_{PL}^{HT}(s) = A\,\tanh\{\alpha + \beta \ln (s/s_0) + \gamma \ln^{\delta} (s/s_0)\};
\label{htpl}
\end{eqnarray}

\item[-]
Hyperbolic Tangent ($HT$) with variant Logarithmic-Law ($LL$)

\begin{eqnarray}
X_{LL}^{HT}(s) = A\,\tanh\{\alpha + \beta \ln (s/s_0) + \gamma \ln \ln (s/s_0)\};
\label{htll}
\end{eqnarray}

\end{description}
\noindent
with the condition
\begin{eqnarray}
s > s_0.
\nonumber
\end{eqnarray}

At last, and still as a matter of a short notation for discussion and data reductions, 
in what follows we shall refer to the sigmoid functions as $logistic$ or $tanh$
and to the variants as $PL$ or $LL$.

\subsection{The Energy Scale}

In principle, the energy scale $s_0$ could be considered as a free fit parameter.
However, that introduces additional nonlinearity in the data reductions and
additional correlations among all the adjustable parameters, which are not easy to
control. In this respect, our strategy was to 
fix the scale, but simultaneously, to investigate its effect in the fit results. 
Specifically, as summarized in \ref{sab}, in our previous analysis we have assumed either $s_0 = 1$ GeV$^2$ 
(an usual choice in phenomenology) \cite{fm12,fm13} or  $s_0 = 25$ GeV$^2$ (the energy
cutoff) \cite{fms15a,fms15b}. Here, we  consider
the reasonable and efficient choice of $s_0$ as the energy threshold for the
scattering states (above the resonance region), namely
\begin{eqnarray}
s_0 = 4 m_p^2,
\nonumber
\end{eqnarray}
where $m_p$ is the proton mass (see also \cite{ms13a}, section 4.2, for further
discussion on this energy scale). We will show (Section 5.5.1) that the selected results do not
depend on the aforementioned choices.

\subsection{Constrained and Unconstrained Fits}

In applying parameterizations (\ref{lpl}), (\ref{lll}), (\ref{htpl}) and (\ref{htll}) to data 
reductions there are two possibilities to treat the asymptotic parameter $A$: 

\begin{description}

\item{(1)} to fix it to an assumed value (as 
one of those displayed in Section 2.2), imposing therefore an asymptotic scenario;

\item{(2)} to treat $A$ as a free parameter, leading to the selection of an asymptotic
scenario. 

\end{description}

In the former case, which we shall refer as
\textit{constrained fit} ($A$ fixed), there are only four free parameters in the case
of the variant $PL$ and only three with the $LL$. In the latter case,
refereed to as \textit{unconstrained fit} ($A$ free) we have five free parameters
with variants $PL$ and four in the $LL$ case.
Certainly, that represents an economical number of free parameters: note
that individual fits to $\sigma_{tot}$ and $\sigma_{el}$, as those in Eqs. (\ref{comp}) and (\ref{tot}),
demand 10 or more parameters for the ratio $X$.

As commented before, our previous analysis have been developed only with the
sigmoid tanh and particular cases of the variant $PL$.
For future discussion these results are summarized in \ref{sab}.
In what follows we  present the fit procedures and all the 
fit results, which will be discussed, in detail, in Section \ref{s5}.

\section{Fit Procedures and Results}
\label{s4}

\subsection{Fit Procedures}

The data reductions have been performed with the objects of the class TMinuit of 
ROOT Framework, with confidence
level fixed at 68 \% \cite{root}.
We have employed the default MINUIT error analysis
with the criteria of full convergence \cite{minuit}.
The error matrix provides the variances
and covariances associated with each free parameter, 
which are used in the analytic evaluation of the uncertainty regions 
associated with fitted and predicted
quantities (through standard error propagation procedures \cite{bev}).

As tests of goodness of fit we shall consider the reduced chi-squared,
$\chi^2/\nu$, where $\nu$ is the number of degrees of freedom and the corresponding 
integrated probability, $P(\chi^2, \nu)$ \cite{bev}.
However, the presence of both statistic and systematic uncertainties in the
experimental data puts some limitation in a formal interpretation of these
tests. For that reason, our goal is not to compare or select fit procedures
or fit results but only to check the statistical consistency of each data
reduction in a rather quantitative way.

Since the parametrization is non-linear in at least three parameters, different initial 
values have been tested
in order to check the stability of the result. In this respect we developed the
procedure that follows.

For each sigmoid (logistic or tanh) and each variant ($PL$ or $LL$),
Eqs. (\ref{lpl}) to (\ref{htll}), 
we first develop the constrained fits
($A$ fixed). In this case we consider the five numerical values displayed in Section 2.2 
as representative of the three scenarios investigated, as summarized
in the following scheme:

\noindent
- above the black disk $\rightarrow$ either $A=1$ (maximum unitarity) or
$A=0.75$ (a possible ``formal" result);

\vspace{0.1cm}
\noindent
- the black disk $\rightarrow$ $A = 0.5$;

\vspace{0.1cm}

\noindent
- below the black disk $\rightarrow$ either $A = 0.436$ (the result from the TOTEM and COMPETE parameterizations, 
Eqs. (\ref{comp}) and (\ref{tot})) or $A = 0.3$ (lowest value we have obtained in 
\cite{fms13,ms13a,ms13b}).

\vspace{0.2cm}

For each fixed $A$, different initial values have been tested for the other parameters 
(4 with $PL$ and 3 with $LL$), until reaching stable convergence and consistent statistical results.

In a second step, using the values of the parameters from the constrained fit result
as initial values (feedback), we have developed the unconstrained fits, with each $A$ 
now as free parameter too.

As already stated, by fixing $A$  we \textit{impose} an asymptotic limit and by
letting $A$ as a free fit parameter, we  \textit{select} an asymptotic
scenario. The following diagram summarizes the cases investigated:
$$
\mbox{sigmoid}\ \left\{\begin{array}{l}
\mbox{logistic}\\
\mbox{tanh}
\end{array}\right.
\quad
\Rightarrow
\quad
\mbox{variant}\ \left\{\begin{array}{l}
PL\\
LL
\end{array}\right.
\quad
\Rightarrow
\quad
\mbox{fit}\ \left\{\begin{array}{l}
\mbox{constrained (A fixed)}\\
\mbox{unconstrained (A free)}
\end{array}\right.
$$

In what follows we display  the fit results with the logistic function 
followed by those with the tanh.
All these results will be discussed in Section \ref{s5}.

\subsection{Fit Results with the Logistic}

\subsubsection{Variant PL}

\begin{figure}[h!]
\centering
\epsfig{file=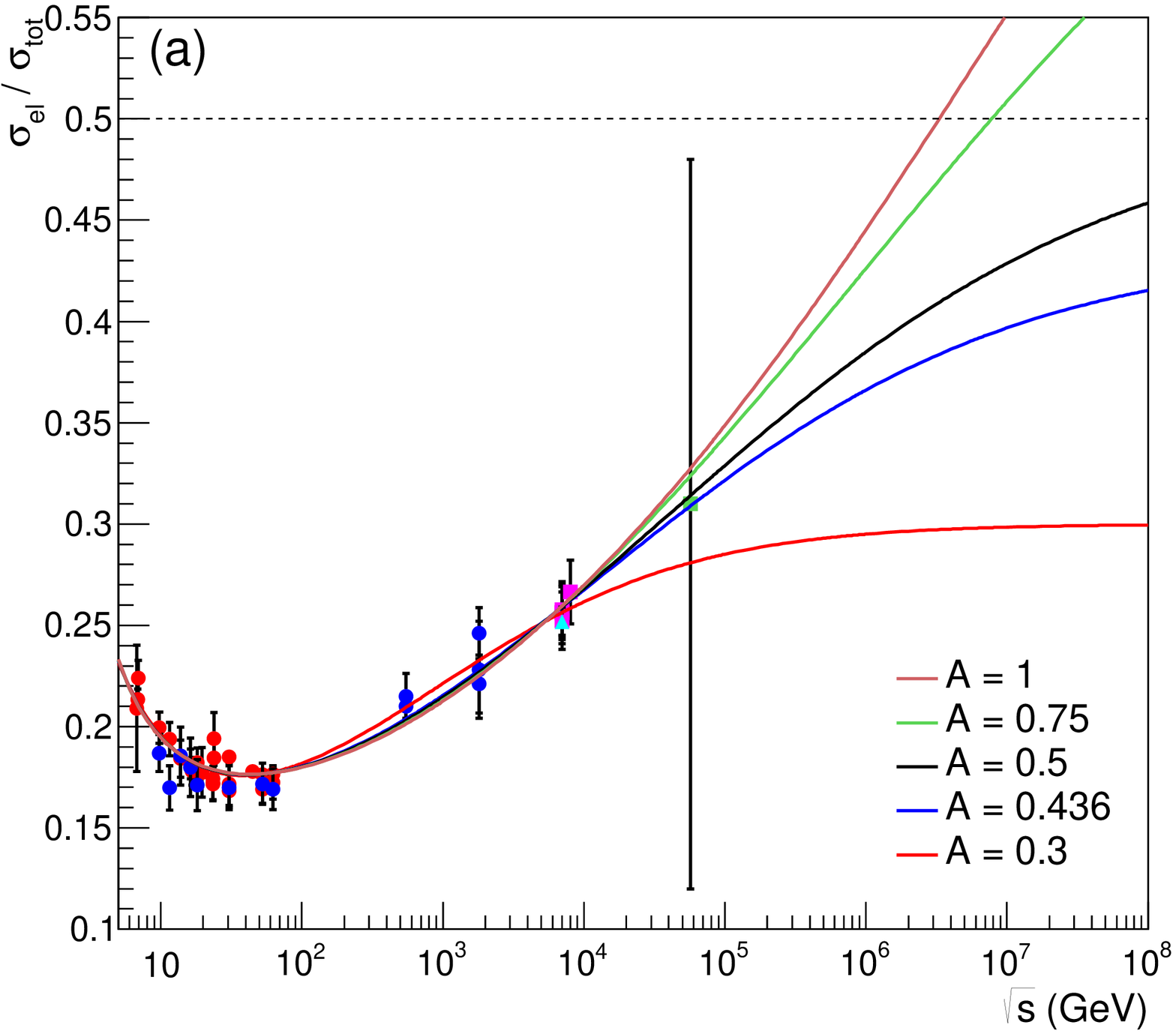,width=8cm,height=7cm}
\epsfig{file=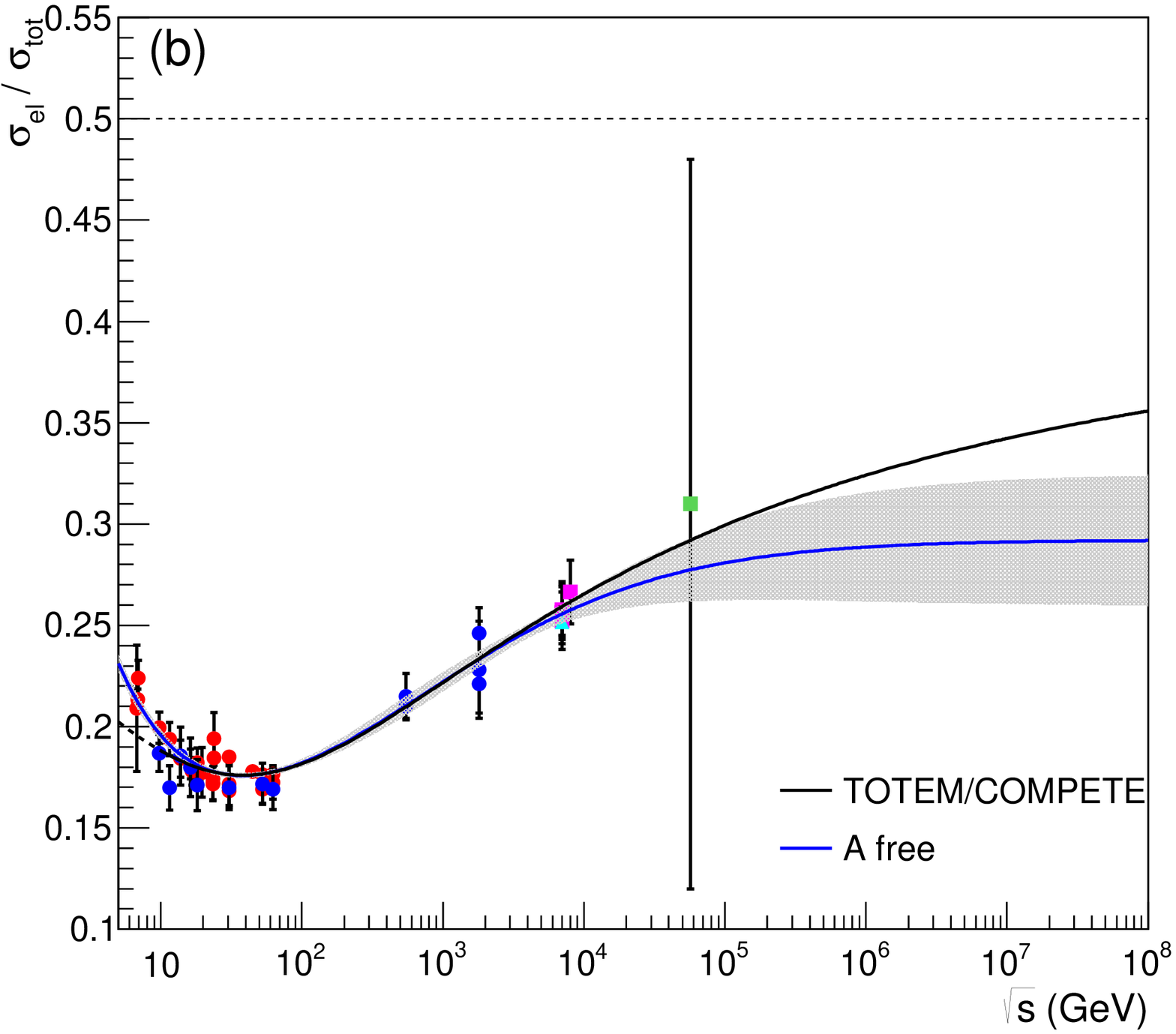,width=8cm,height=7cm}
\caption{Fit results with the logistic function, variant $PL$ (Eq. (\ref{lpl})): (a) constrained case ($A$ fixed, Table \ref{t2}); (b) unconstrained case ($A$ free, Table \ref{t3}) with the corresponding
uncertainty region and the result from the TOTEM and COMPETE parameterizations, Eqs. (\ref{comp}) and (\ref{tot}).
Legend for data given in Figure \ref{f1}.}
\label{f2}
\end{figure}

In the case of the constrained fit, for each fixed $A$ we have four fit parameters. 
The results and statistical information on each fit are displayed in Table \ref{t2} 
and the corresponding curves, compared with the experimental data, 
in Fig. \ref{f2}(a). For clarity, the plotted curves correspond only to the
central values of the free parameters (i.e. without the uncertainty regions).

With these results for the parameters as initial values, including
each $A$ value, the unconstrained fits have been developed. The results
are displayed in Table \ref{t3} 
and show that all data reductions
indicate almost the same goodness of fit ($\chi^2/\nu$ and $P(\chi^2,\nu)$) and the
\textit{same asymptotic central value}, namely $A = 0.292$.
Although, in each case, the values of the parameters $\alpha$, $\beta$, $\gamma$
and $\delta$ 
may differ, once plotted together all curves corresponding
to the central values of the parameters overlap.
That is shown in Fig.
\ref{f2}(b), where the corresponding uncertainty region has been
evaluated through error propagation from the fit parameters (Table \ref{t3})
within one standard deviation.
For comparison, we have also displayed the ratio $X(s)$ obtained
through the TOTEM and COMPETE parameterizations, Eqs. (\ref{comp}) and (\ref{tot}).

\begin{table}[ht]
\centering
\caption{Fit results with the logistic function, variant $PL$ (Eq. (\ref{lpl})) and constrained case 
($A$ fixed), $\nu = 38$ (Fig. \ref{f2}(a)).} 
\vspace{0.1cm}
\begin{tabular}{c c c c c c c}\hline
$A$   &       $\alpha$      &        $\beta$       &       $\gamma$       & $\delta$ & $\chi^2/\nu$ & $P(\chi^2,\nu)$\\
fixed &                     &                      &                      &   &                    &                 \\\hline
0.3   & 125.5(1.5)  & 0.328(22)  & -123.7(1.5)  & 1.56(12)$\times 10^{-2}$  & 0.811 & 0.790 \\
0.436 & 169.12(10)  & 0.1828(72) & -168.59(10)  & 6.62(30)$\times 10^{-3}$  & 0.882 & 0.677 \\
0.5   & 211.37(96)  & 0.1627(63) & -211.156(95) & 4.74(21)$\times 10^{-3}$  & 0.899 & 0.647 \\
0.75  & 69.517(80)  & 0.1315(51) & -70.027(79)  & 1.148(52)$\times 10^{-2}$ & 0.932 & 0.589 \\
1.0   & 82.898(75)  & 0.1195(46) & -83.825(73)  & 8.78(40)$\times 10^{-3}$  & 0.944 & 0.568 \\\hline
\end{tabular}
\label{t2}
\end{table}
\begin{table}[ht]
\centering
\caption{Fit results with the logistic function, variant $PL$ (Eq. (\ref{lpl}))
and unconstrained case ($A$ free), $\nu=37$ (Fig. \ref{f2}(b)).}
\vspace{0.1cm}
\begin{tabular}{c c c c c c c c}\hline
$A$     & $A$               &     $\alpha$     &     $\beta$      &     $\gamma$      & $\delta$ & $\chi^2/\nu$ & $P(\chi^2,\nu)$\\
initial & free              &                  &                  &                   &          &             &                 \\\hline
0.3     & 0.292(33) & 125.6(1.5) & 0.35(11) & -123.6(1.5) & 1.66(51)$\times 10^{-2}$ & 0.831 & 0.756 \\
0.436   & 0.292(33) & 169.9(1.3) & 0.35(11) & -167.9(1.3) & 1.22(38)$\times 10^{-2}$ & 0.831 & 0.756 \\
0.5     & 0.292(33) & 212.2(1.2) & 0.35(11) & -210.2(1.2) & 9.8(3.0)$\times 10^{-3}$ & 0.831 & 0.757 \\
0.75    & 0.292(32) & 70.5(1.9)  & 0.36(12) & -68.5(1.9)  & 2.98(91)$\times 10^{-2}$ & 0.832 & 0.755 \\
1.0     & 0.292(32) & 70.5(1.9)  & 0.36(12) & -68.5(1.9)  & 2.98(91)$\times 10^{-2}$ & 0.832 & 0.755 \\\hline
\end{tabular}
\label{t3}
\end{table}

\subsubsection{Variant LL}

\begin{figure}[h!]
\centering
\epsfig{file=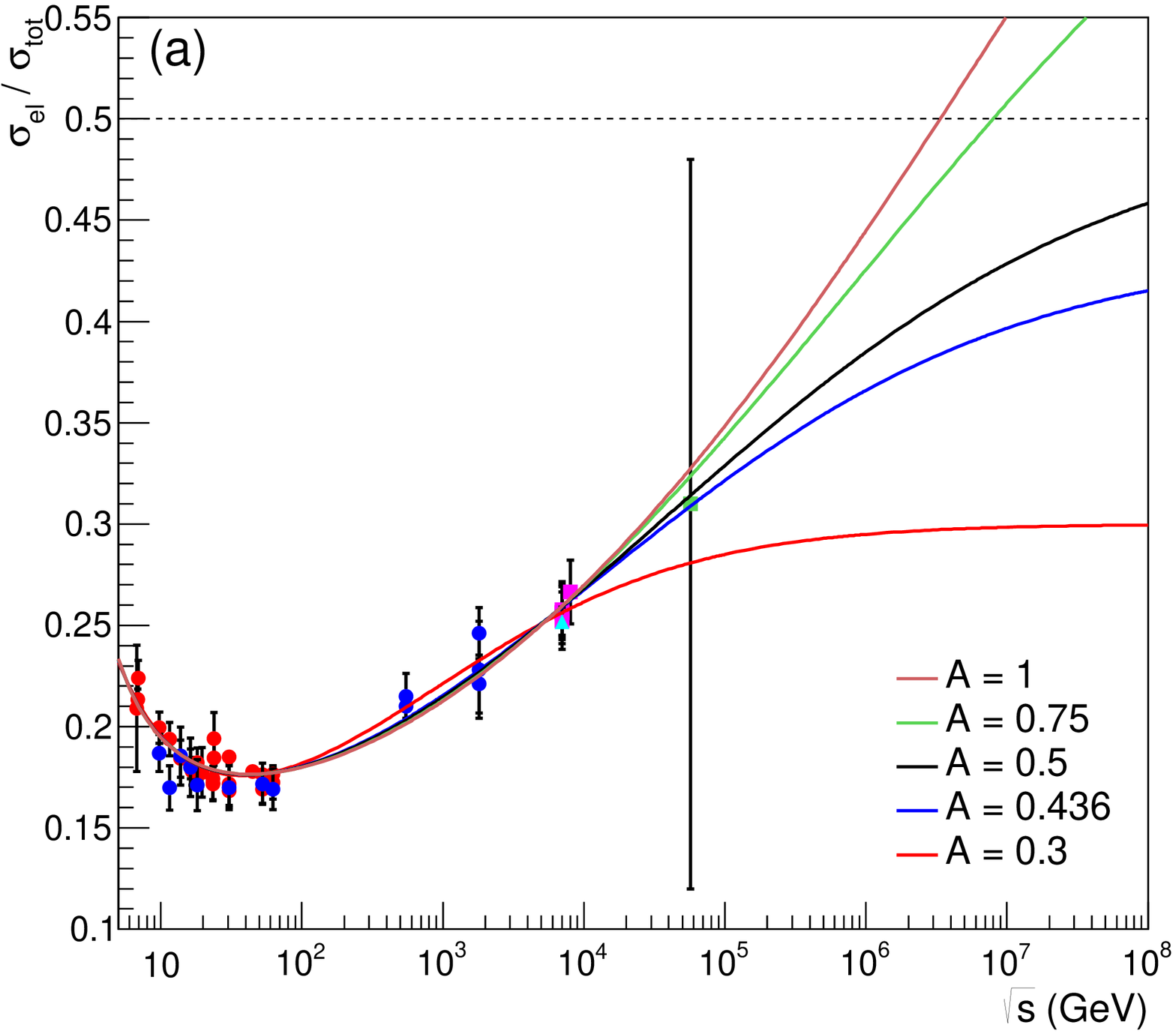,width=8cm,height=7cm}
\epsfig{file=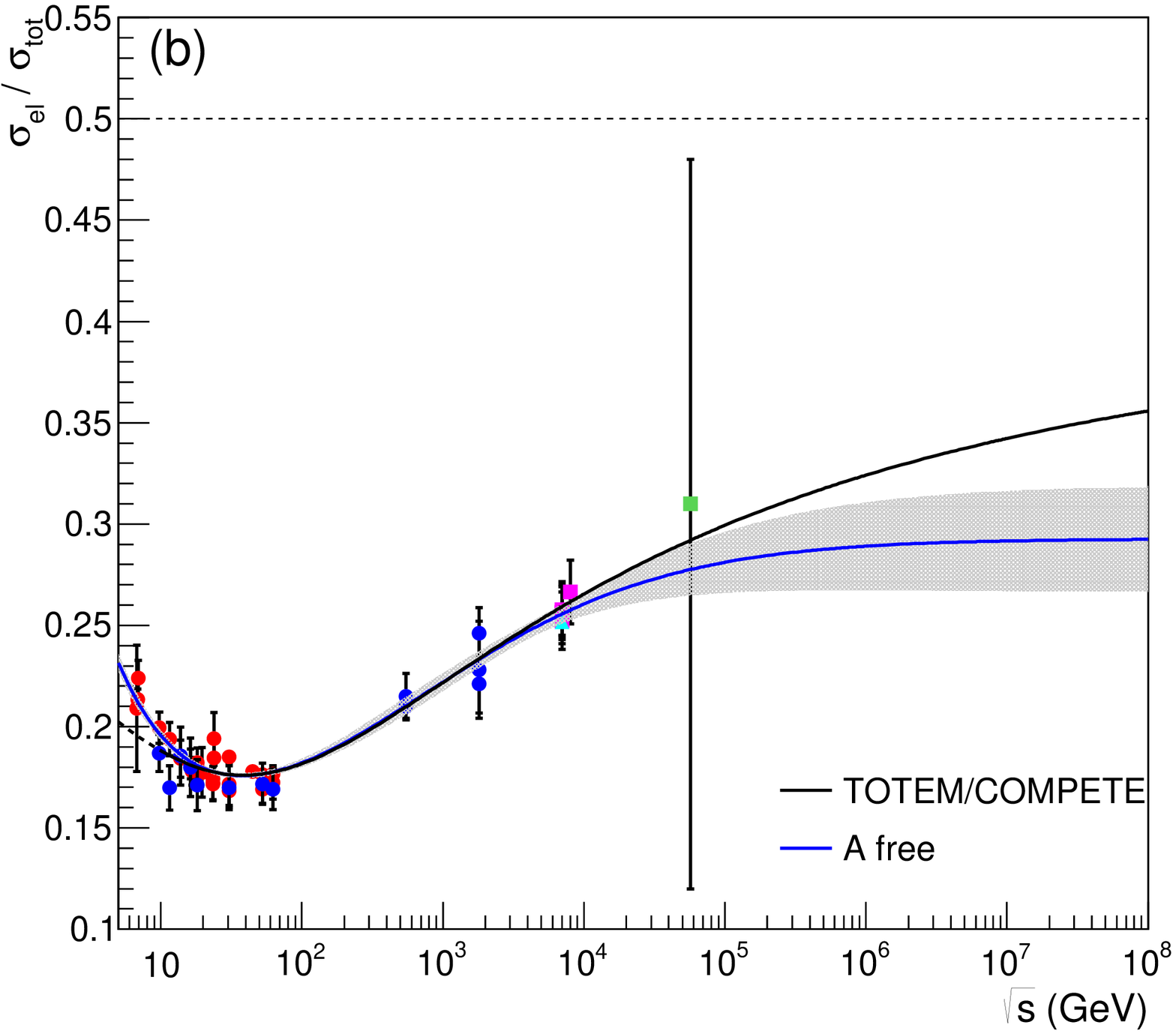,width=8cm,height=7cm}
\caption{Fit results with the logistic function, variant $LL$ (Eq. (\ref{lll})): 
(a) constrained case ($A$ fixed, Table \ref{t4}); 
(b) unconstrained case ($A$ free, Table \ref{t5}) with the corresponding
uncertainty region and the result from the TOTEM and COMPETE parameterizations, Eqs. (\ref{comp}) 
and (\ref{tot}). Legend for data given in Figure \ref{f1}.}
\label{f3}
\end{figure}

The same procedure has been developed with the variant $LL$. In this case we have only
three parameters ($A$ fixed) or four ($A$ free). The constrained fit results
are displayed in Table \ref{t4} and Figure \ref{f3}(a) and the unconstrained
case in Table \ref{t5}.
Once more, all fit results converged to the same solution
in statistical grounds and in the values of all the parameters. The corresponding
curve including the uncertainty region is shown in Figure \ref{f3}(b).

\begin{table}[ht]
\centering
\caption{Fit results with the logistic function, variant $LL$ (Eq. (\ref{lll})) and constrained case 
($A$ fixed), $\nu = 39$ (Fig. \ref{f3}(a)).} 
\vspace{0.1cm}
\begin{tabular}{c c c c c c}\hline
$A$   &       $\alpha$      &        $\beta$       &       $\gamma$       &  $\chi^2/\nu$ & $P(\chi^2,\nu)$\\
fixed &                     &                      &   &                    &                 \\\hline
0.3   & 1.90(13)   & 0.324(22)  & -1.95(14)  &  0.789 & 0.824 \\
0.436 & 0.534(79)  & 0.182(10)  & -1.122(78) &  0.858 & 0.720 \\
0.5   & 0.221(71)  & 0.1620(91) & -1.005(69) &  0.875 & 0.692 \\
0.75  & -0.503(58) & 0.1302(71) & -0.813(56) &  0.907 & 0.638 \\
1.0   & -0.922(53) & 0.1186(65) & -0.742(51) &  0.918 & 0.617 \\\hline
\end{tabular}
\label{t4}
\end{table}
\begin{table}[ht]
\centering
\caption{Fit results with the logistic function, variant $LL$ (Eq. (\ref{lll})) and unconstrained case ($A$ free),
$\nu=38$ (Fig. \ref{f3}(b)).}
\vspace{0.1cm}
\begin{tabular}{c c c c c c c}\hline
$A$     & $A$               &     $\alpha$     &     $\beta$      &     $\gamma$      & $\chi^2/\nu$ & $P(\chi^2,\nu)$\\
initial & free              &                  &                  &          &             &                 \\\hline
0.3     & 0.293(26) & 2.05(59) & 0.346(87) & -2.07(50) & 0.808 & 0.794 \\
0.436   & 0.293(26) & 2.05(59) & 0.346(88) & -2.07(50) & 0.808 & 0.794 \\
0.5     & 0.293(26) & 2.05(59) & 0.346(88) & -2.07(50) & 0.808 & 0.794 \\
0.75    & 0.293(26) & 2.05(59) & 0.346(88) & -2.07(50) & 0.808 & 0.794 \\
1.0     & 0.293(26) & 2.05(59) & 0.346(88) & -2.07(50) & 0.808 & 0.794 \\\hline
\end{tabular}
\label{t5}
\end{table}

\subsection{Fit Results with the Hyperbolic Tangent}
\label{sac}

The same procedures discussed for the logistic function
have been applied in the case of the tanh, as follows.

\subsubsection{Variant PL}

\begin{figure}[h!]
\centering
\epsfig{file=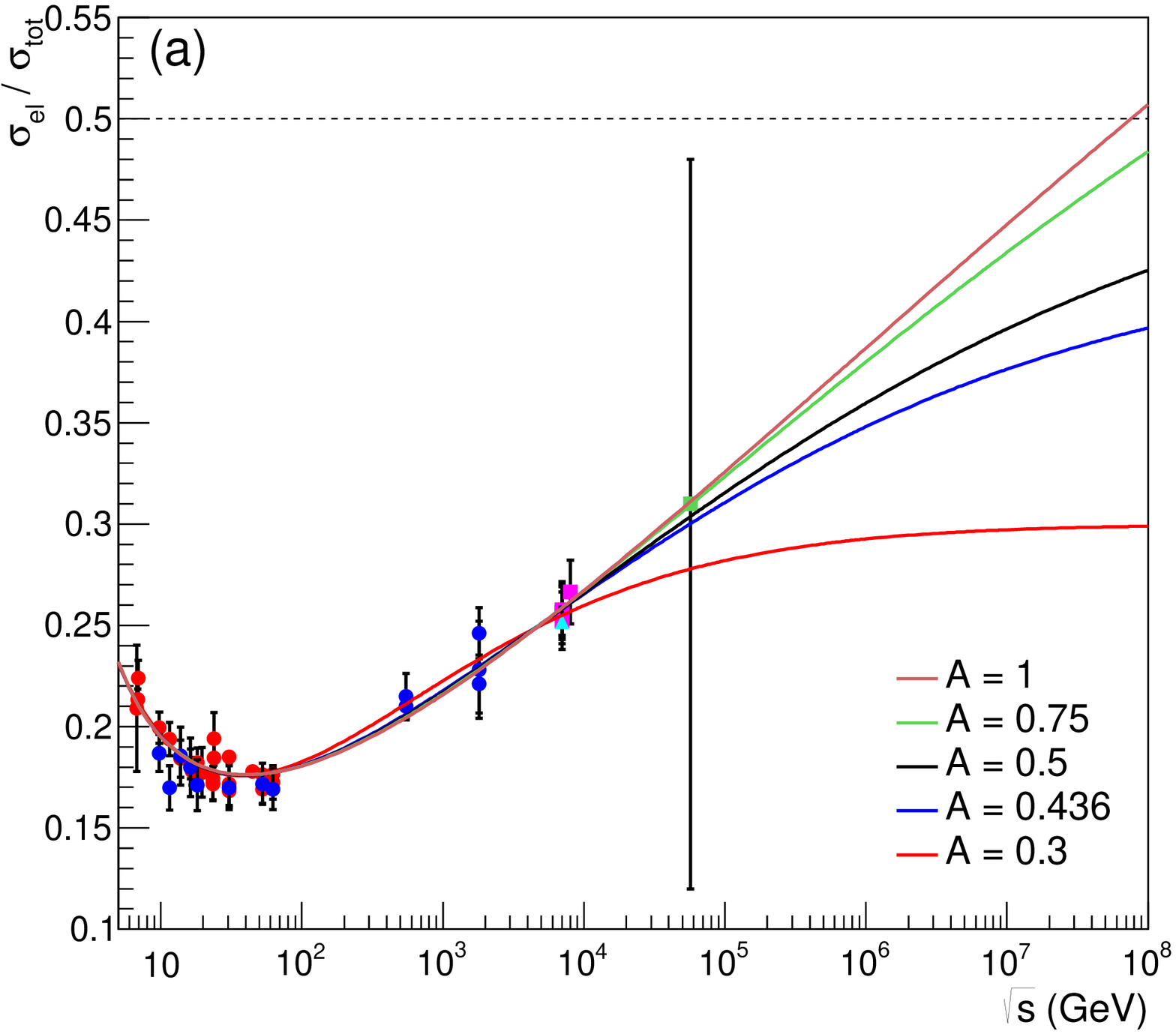,width=8cm,height=7cm}
\epsfig{file=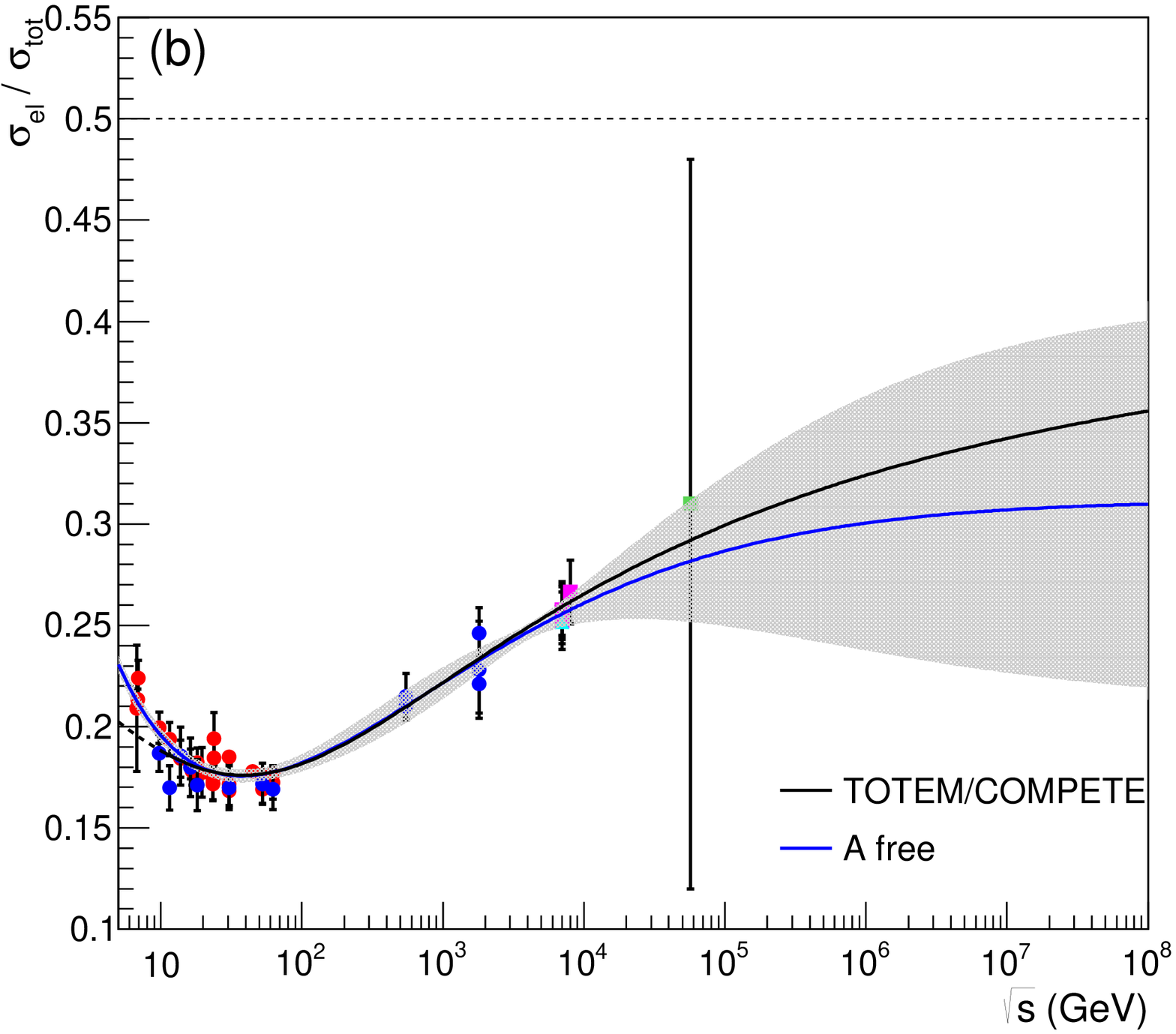,width=8cm,height=7cm}
\caption{Fit results with the tanh, variant $PL$ (Eq. (\ref{htpl})): 
(a) constrained case ($A$ fixed, Table
\ref{t6}); (b) unconstrained case ($A$ free, Table \ref{t7}) with the corresponding
uncertainty region and the result from the TOTEM and COMPETE parameterizations, Eqs. (\ref{comp}) and (\ref{tot}).
Legend for data given in Figure \ref{f1}.}
\label{f4}
\end{figure}

The fit results with the variant $PL$ are displayed in Table \ref{t6} (constrained
case), Table \ref{t7} (unconstrained case) and Fig. \ref{f4}.
 
\begin{table}[ht]
\centering
\caption{Fit results with the tanh, variant $PL$ (Eq. (\ref{htpl})) and 
constrained case ($A$ fixed), $\nu = 38$ (Fig. \ref{f4}(a)).} 
\vspace{0.1cm}
\begin{tabular}{c c c c c c c}\hline
$A$   &       $\alpha$      &        $\beta$       &       $\gamma$       & $\delta$ & $\chi^2/\nu$ & $P(\chi^2,\nu)$\\
fixed &                     &                      &                      &   &                    &                         \\\hline
0.3   & 125.23(24)   & 0.13243(93) & -123.95(24) & 6.31(48)$\times 10^{-3}$  &   0.818          & 0.780  \\
0.436 & 169.21(13)   & 0.0588(35)  & -168.50(13) & 2.23(15)$\times 10^{-3}$  &   0.844          & 0.740 \\
0.5   & 211.44(10)   & 0.0477(28)  & -210.83(10) & 1.380(96)$\times 10^{-3}$ &   0.853          & 0.724 \\
0.75  & 68.06(14)    & 0.0283(16)  & -67.67(14)  & 2.56(18)$\times 10^{-3}$  &   0.870          & 0.697 \\
1.0   & 83.50(10)    & 0.0204(16)  & -83.22(10)  & 1.51(10)$\times 10^{-3}$  &   0.875          & 0.687 \\\hline
\end{tabular}
\label{t6}
\end{table}
\begin{table}[ht]
\centering
\caption{Fit results with the tanh, variant $PL$ (Eq. (\ref{htpl})) and 
unconstrained case ($A$ free), $\nu=37$ (Fig. \ref{f4}(b)).} 
\vspace{0.1cm}
\begin{tabular}{c c c c c c c c}\hline
$A$     & $A$               &     $\alpha$    &     $\beta$      &     $\gamma$      & $\delta$ & $\chi^2/\nu$ & $P(\chi^2,\nu)$\\
initial & free              &                 &                  &                   &          &             &                         \\\hline
0.3     & 0.31(10)   & 125.18(43) & 0.12(11)  & -123.99(43) & 5.7(4.9)$\times 10^{-3}$   & 0.838    & 0.746 \\
0.436   & 0.31(10)   & 169.45(43) & 0.12(11)  & -168.25(42) & 4.2(3.8)$\times 10^{-3}$   & 0.838    & 0.746 \\ 
0.5     & 0.312(49)  & 211.73(18) & 0.118(52) & -210.54(17) & 3.3(1.4)$\times 10^{-3}$   & 0.837    & 0.746 \\
0.75    & 0.31(10)   & 68.42(48)  & 0.12(11)  & -67.23(48)  & 1.04(0.93)$\times 10^{-2}$ & 0.838    & 0.745 \\
1.0     & 0.31(10)   & 83.92(46)  & 0.12(11)  & -82.73(45)  & 8.5(7.4)$\times 10^{-3}$   & 0.838    & 0.746 \\\hline
\end{tabular}
\label{t7}
\end{table}

\subsubsection{Variant LL}

\begin{figure}[h!]
\centering
\epsfig{file=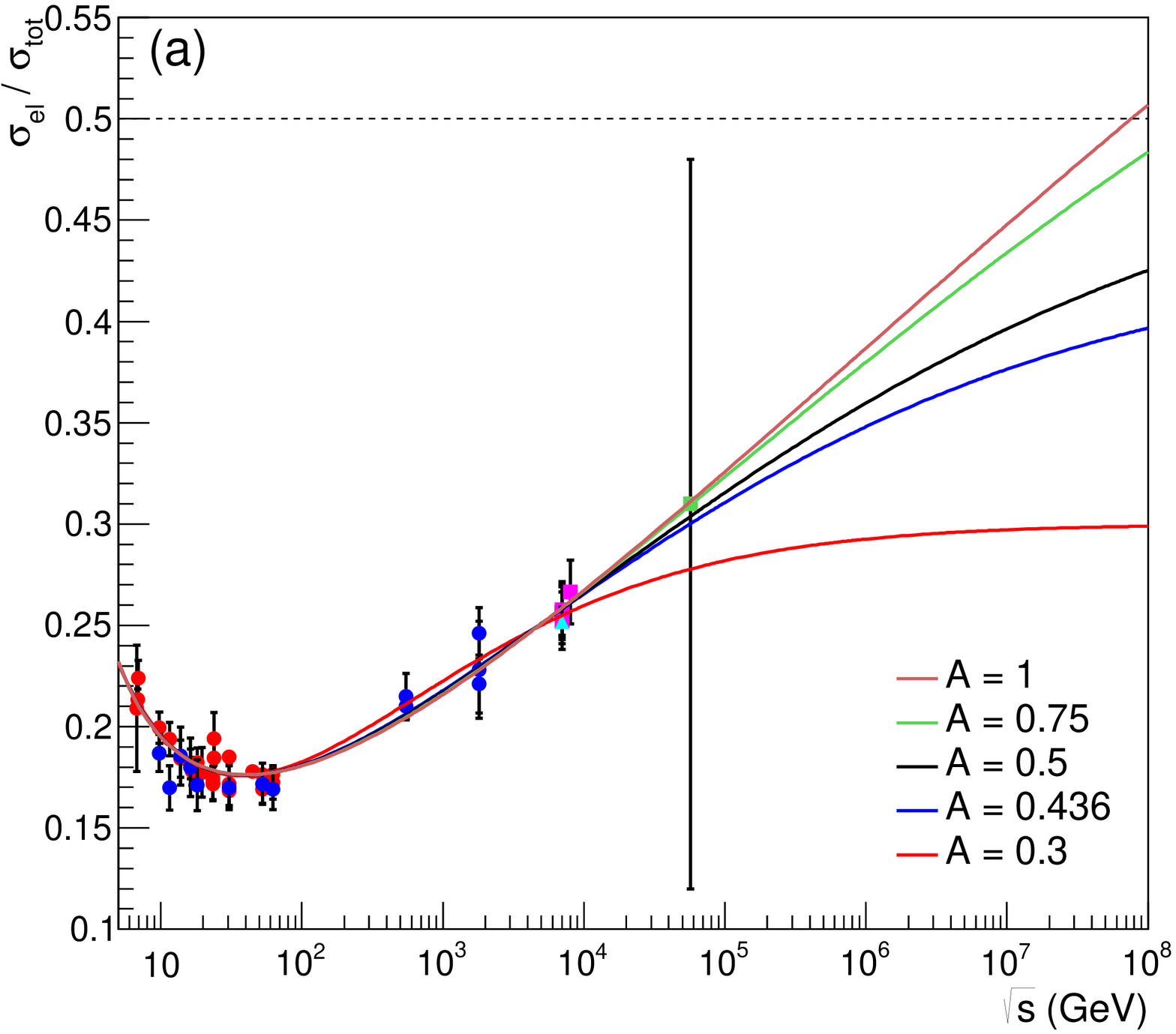,width=8cm,height=7cm}
\epsfig{file=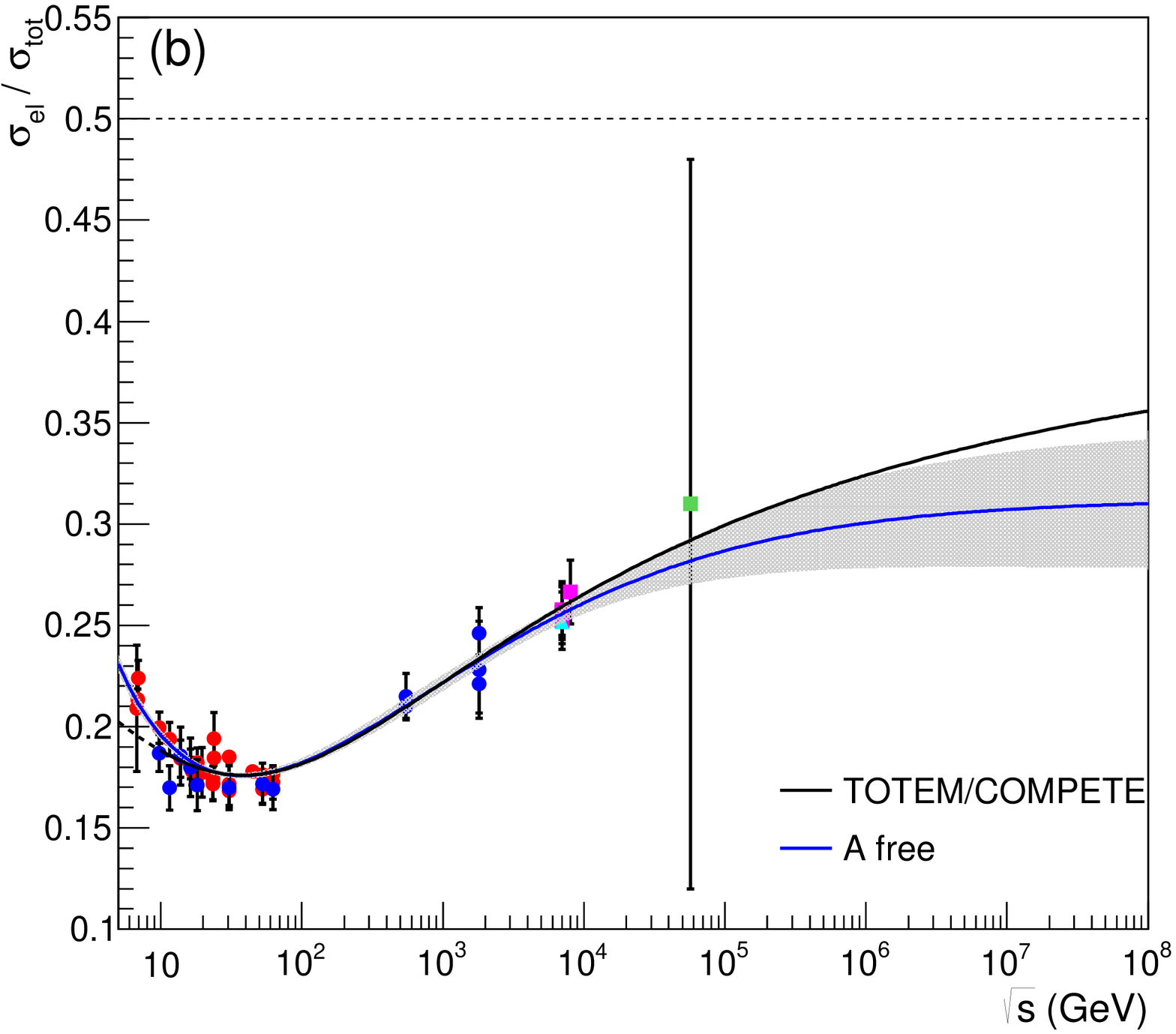,width=8cm,height=7cm}
\caption{Fit results with the tanh, variant $LL$ (Eq. (\ref{htll})): 
(a) constrained case ($A$ fixed, Table
\ref{t8}); (b) unconstrained case ($A$ free, Table \ref{t9}) with the corresponding
uncertainty region and the result from the TOTEM and COMPETE parameterizations, Eqs. (\ref{comp}) and (\ref{tot}).
Legend for data given in Figure \ref{f1}.}
\label{f5}
\end{figure}

The fit results with the variant $LL$ are displayed in Table \ref{t8} (constrained
case), Table \ref{t9} (unconstrained case) and Fig. \ref{f5}.

\begin{table}[ht]
\centering
\caption{Fit results with the tanh, variant $LL$ (Eq. (\ref{htll})) and constrained case ($A$ fixed),
$\nu = 39$ (Fig. \ref{f5}(a)).} 
\vspace{0.1cm}
\begin{tabular}{c c c c c c}\hline
$A$   &       $\alpha$      &        $\beta$     &       $\gamma$       & $\chi^2/\nu$ & $P(\chi^2,\nu)$\\
fixed &                     &                    &                      &                    &                         \\
\hline
0.3   & 1.289(54)   & 0.1317(90)  & -0.786(58)   &   0.796          & 0.813  \\
0.436 & 0.718(25)   & 0.0587(35)  & -0.358(25)   &   0.822          & 0.777 \\
0.5   & 0.605(20)   & 0.0476(28)  & -0.291(20)   &   0.831          & 0.763 \\
0.75  & 0.381(12)   & 0.0283(16)  & -0.174(12)   &   0.847          & 0.737 \\
1.0   & 0.2811(89)  & 0.0204(11)  & -0.1259(87)  &   0.853          & 0.729 \\\hline
\end{tabular}
\label{t8}
\end{table}
\begin{table}[ht]
\centering
\caption{Fit results with the tanh, variant $LL$ (Eq. (\ref{htll})) and unconstrained
case ($A$ free), $\nu=38$ (Fig. \ref{f5}(b)).} 
\vspace{0.1cm}
\begin{tabular}{c c c c c c c}\hline
$A$     & $A$               &     $\alpha$    &     $\beta$      &     $\gamma$    & $\chi^2/\nu$ & 
$P(\chi^2,\nu)$\\ 
initial & free & & & & & \\
\hline
0.3     & 0.312(35)   & 1.19(25) & 0.117(37)  & -0.70(21) & 0.815    & 0.783 \\
0.436   & 0.312(48)   & 1.19(35) & 0.117(51)  & -0.70(29) & 0.815    & 0.783 \\
0.5     & 0.312(35)   & 1.19(25) & 0.117(36)  & -0.70(21) & 0.815    & 0.783 \\
0.75    & 0.312(35)   & 1.19(25) & 0.117(37)  & -0.70(21) & 0.815    & 0.783 \\
1.0     & 0.312(35)   & 1.19(25) & 0.117(37)  & -0.70(21) & 0.815    & 0.783 \\\hline
\end{tabular}
\label{t9}
\end{table}

\section{Discussion and Conclusions on the Fit Results}
\label{s5}

In this section we treat the fit results obtained
with both the logistic (Section 4.2) and the tanh (Section 4.3).
In the first three subsections, we summarize and discuss in a comparative way 
the data reductions with:
constrained/unconstrained fits (Section 5.1), logistic/tanh (Section 5.2) 
and variants $PL$/$LL$ (Section 5.3); after that we treat the inflection point 
(Section 5.4), present our 
conclusions on the fit results (Section 5.5), which will led us to the selection of 
two asymptotic scenarios: semi-transparent and black limits.

\subsection{Constrained and Unconstrained Fits}

In the case of \textit{constrained} fits ($A$ fixed),
the values of the parameters, statistical information and curves are presented 
in Table \ref{t2}, Fig. \ref{f2}(a) (logistic - $PL$), Table \ref{t4}, Fig. \ref{f3}(a) (logistic - $LL$), 
Table \ref{t6}, Fig. \ref{f4}(a) (tanh - $PL$) and Table \ref{t8}, Fig. \ref{f5}(a) (tanh - $LL$).
As already commented, for clarity, the curves in the figures correspond to the central values 
of the parameters.
However, error propagation from the fit parameters (one standard deviation) lead to typical uncertainty 
regions as those displayed in part (b) of each figure (to be discussed later).

From the Tables, for $\nu$ = 38 or 39, the values of the $\chi^2/\nu$ lie in a 
typical interval 0.79 - 0.94 and those of the $P(\chi^2, \nu)$ in the corresponding
interval 0.82 - 0.57, indicating, therefore, statistically consistent fit results
in all cases investigated. We also note that the corresponding curves are consistent
with all the experimental data analyzed (mainly within the uncertainty regions,
not shown in the figures).

In the case of \textit{unconstrained} fits ($A$ free),
the value of the parameters, statistical information and curves are presented 
in Table \ref{t3}, Fig. \ref{f2}(b) (logistic - $PL$), Table \ref{t5}, Fig. \ref{f3}(b) (logistic - $LL$), 
Table \ref{t7}, Fig. \ref{t4}(b) (tanh - $PL$), Table \ref{t9}, Fig. \ref{f5}(b) (tanh - $LL$).

Here we note a remarkable fact: all fit results converged to practically the
same solution, specially in what concerns the asymptotic limit. These values 
are summarized in Table \ref{t10}. Small differences in the value 
of the other fit parameters are discussed in what follows.

\begin{table}[h!]
\centering
\caption{Summary of the asymptotic limits obtained in the unconstrained fits,
($A$ free) with the logistic, tanh, variants $PL$, $LL$ and
the corresponding $\nu$ and $\chi^2/\nu$. It is also shown $\nu$ and $\chi^2/\nu$
in the case of the black disk ($A$ = 0.5).} 
\vspace{0.1cm}
\begin{tabular}{c c c c c}
\hline
sigmoid: &\multicolumn{2}{c}{logistic}&\multicolumn{2}{c}{tanh} \\
variant: &    $PL$       &   $LL$     & $PL$      &   $LL$         \\
\hline
$A$ free        &   0.292(33)   & 0.293(26)  & 0.31(10)  & 0.312(48)       \\
$\nu$           &   37          & 38         & 37        & 38              \\
$\chi^2/\nu$    &   0.832       & 0.808      & 0.838     & 0.815           \\ 
\hline                 
$A$ fixed       &  0.5          & 0.5        & 0.5       & 0.5\\
$\nu$           &  38           & 39         & 38        & 39              \\
$\chi^2/\nu$    &  0.899        & 0.875      & 0.853     & 0.831           \\ 
\hline 
\end{tabular}
\label{t10}
\end{table}

\subsection{Sigmoid Functions: Logistic and Hyperbolic Tangent}

First, let us compare the \textit{constrained} results in part (a) of
Figs. \ref{f2} and \ref{f3} (logistic) with those in part (a) of Figs. \ref{f4} and \ref{f5}
(tanh). We note that, above the experimental data,  the rise of $X(s)$ 
with the logistic is faster than with the tanh and the differences
increase as $A$ increases (from 0.3 up to 1.0). For example, in the case
of $A$ = 1 (fixed), at $\sqrt{s}$ = 10$^7$ GeV, $X \approx$ 0.55 with the logistic
and $X \approx$ 0.45 with the tanh in both variants, $PL$ and $LL$.

On the other hand, in the \textit{unconstrained} cases (part (b) of the same figures),
these differences are negligible, even in the asymptotic region. The small differences
in the central values of $A$ (namely $\approx$ 0.29 for the logistic and $\approx$
0.31 for the tanh) are also negligible within the uncertainties
in these parameters (Table \ref{t10}).

\subsection{Variants $PL$ and $LL$} 

In the \textit{constrained} fits, the values of the free parameters differ
substantially with variants $PL$ and $LL$, in both cases:
logistic (compare Tables \ref{t2} and \ref{t4}) and tanh
(compare Tables \ref{t6} and \ref{t8}). Despite these differences, once plotted
together, all curves overlap,
as can be seen in parts (a) of Fig. \ref{f2} ($PL$) and Fig. \ref{f3} ($LL$) for
the logistic and parts (a) of Fig. \ref{f4} ($PL$) and Fig. \ref{f5} ($LL$) for the tanh.

In the \textit{unconstrained} case and variant $PL$ (Table \ref{t3} for the logistic
and Table \ref{t7} for the tanh) all results lead to practically the same
asymptotic values, $A$ = 0.292 (logistic) and $A$ = 0.31 (tanh),
with some differences in the central values of the other parameters.
On the other hand, with variant $LL$ \textit{the central values of the parameters are
all the same up to three figures} with the logistic (Table \ref{t5}) and also with 
the tanh (Table \ref{t9}). The same is true for the goodness of fit in each case.

\subsection{Inflection Point} 

By construction our parameterizations present a change of curvature (an inflection
point), determined by the root of the second derivative. In what concerns the
ratio between elastic and total cross section, this inflection suggests a change
in the dynamics of the collision process and that will be discussed in Section
7.1. For future reference, we display in Table \ref{t11} the position in the c.m. energy
of the inflection point for all fits developed. We note that this position increases
with the value of the asymptotic ratio $A$, but it is restricted to an interval
$\sqrt{s} \approx$ 80 - 100 GeV. We shall return to this important feature in Section
7.1.

\begin{table}[h!]
\centering
\caption{Inflection point: roots of the second derivative of $X(\sqrt{s})$ in GeV
for each fit developed.} 
\vspace{0.1cm}
\begin{tabular}{c c c c c}
\hline
 &\multicolumn{2}{c}{logistic}&\multicolumn{2}{c}{tanh} \\
   $A$ &    $PL$  &   $LL$  & $PL$  &   $LL$        \\
\hline
free   &   81.6   &  81.2   &  80.1     & 80.0  \\
0.3    &   82.9   &  82.4   &   78.7    & 78.5  \\
0.436  &   92.2   &  92.0   &   86.7    & 86.6  \\ 
0.5    &   93.7   &  93.5   &   87.8    & 87.8  \\
0.75   &   96.5   &  96.1   &   89.7    & 89.6  \\
1.0    &   97.4   &  97.1   &   90.3    & 90.2  \\ 
\hline 
\end{tabular}
\label{t11}
\end{table}

\subsection{Conclusions on the Fit Results}

In what follows, based on the above discussions, we present our conclusions on
the fit results, which will lead us to select one variant ($LL$) and two scenarios 
(semi-transparent and black limits)
for further discussions and studies.

\subsubsection{Main Conclusions}

As regards \textit{constrained} fits, given the relative large uncertainties
in the experimental data and the small differences in the values of
$\chi^2/\nu$ and  $P(\chi^2, \nu)$ for $\nu$ = 38 - 39, 
we understand that all the fit results are statistically consistent 
with the dataset and equally probable on statistical grounds,
even in the extrema cases of $A$ = 0.3 and $A$ = 1. In other words,
despite the large differences in the extrapolated results (at the highest and asymptotic 
energy regions), the constrained fits does not allow to select an asymptotic
scenario.
That leads to an important consequence: although consistent with the
experimental data, the black disk does not represent an unique or definite
solution.

With regards to \textit{unconstrained} fits,
independently of the sigmoid function (logistic or tanh) and variant ($PL$ or $LL$),
the data reductions lead to unique solutions within the uncertainties,
indicating a scenario below the black disk, with central value of $A$
around 0.29 (logistic) and around 0.31 (tanh),
as summarized in Tab. \ref{t10}.
Given the convergent character of these
solutions, we consider \textit{the unconstrained fits as the preferred results of
this analysis}, namely we understand that \textit{the data reductions favor a semi-transparent
(or gray) scenario}.
The small differences among our four asymptotic results may be associated with
a kind of ``uncertainty" in the choices of $S(f)$ and $f(s)$. In this sense,
once constituting independent results (central values and uncertainties),
we can infer a \textit{global} ($g$) asymptotic value, for example, through the arithmetic mean and addition of
the uncertainties in quadrature (Table \ref{t10}):
\begin{eqnarray}
A_g = 0.30 \pm 0.12.
\label{g}
\end{eqnarray}

It is important to notice that, within the uncertainties,
this result is consistent with our previous analysis using the
tanh, $\delta$ = 0.5 (fixed) and  energy scale at 25 GeV$^2$ (the energy
cutoff), namely  $A$ = 0.36(8) in \cite{fms15a} and $A$ = 0.332(49) in \cite{fms15b}.
In other words, \textit{the convergent result does not depend on the energy scale}.
Moreover, within the uncertainties, the above value is in plenty agreement
with the limits obtained through
individual fits to $\sigma_{tot}$ and $\sigma_{el}$ data, using different
variants and procedures (see Fig. 10 in Ref. \cite{ms13b}).

Concerning the variants $PL$ and $LL$, given the goodness of fits, the uniqueness
character of the convergent solutions (represented by the same central values of the
parameters up to three figures) and the smaller
number of free parameters, we select the \textit{variant $LL$ as our most representative result}.
From Tab. \ref{t10}, 
we can also infer another mean value now \textit{restricted} ($r$) to variant $LL$
and the two sigmoid functions:
\begin{eqnarray}
A_r = 0.303 \pm 0.055.
\label{r}
\end{eqnarray}
Within the uncertainties, this result is also in plenty agreement with the aforementioned analyses.
In what follows we shall focus our discussion, predictions and extension to other
quantities to this variant.

\subsubsection{Selected Results and Scenarios}

Although the unconstrained fits have led to an unique semi-transparent (or gray) scenario,
we have shown that the constrained fits with the black disk present also consistent
descriptions of the experimental data. Based on the ubiquity character of 
the black disk limit in the phenomenological context, 
we shall consider also this case as a selected scenario for comparative 
predictions and discussions.

\begin{figure}[h!]
\centering
\epsfig{file=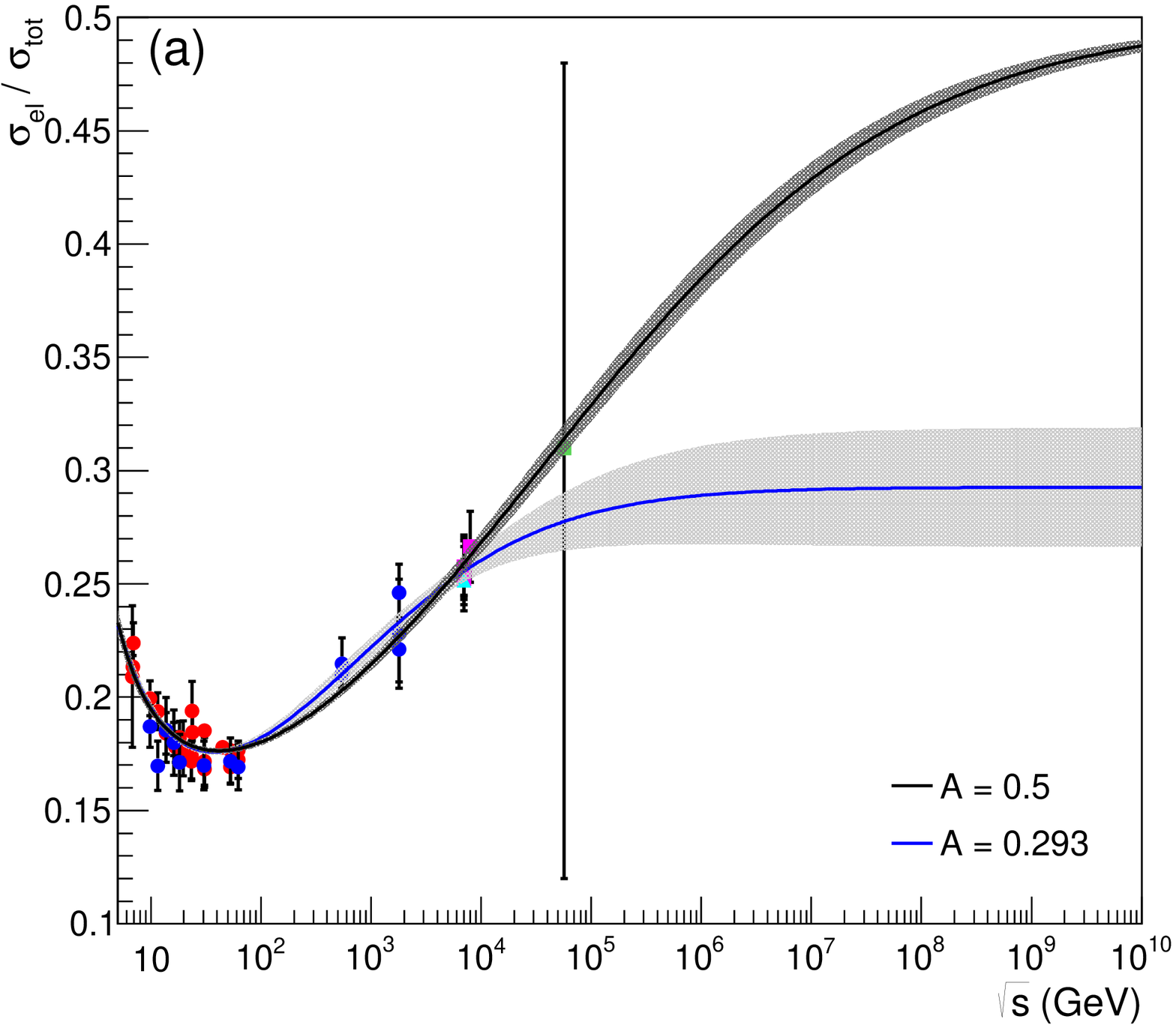,width=8cm,height=7cm}
\epsfig{file=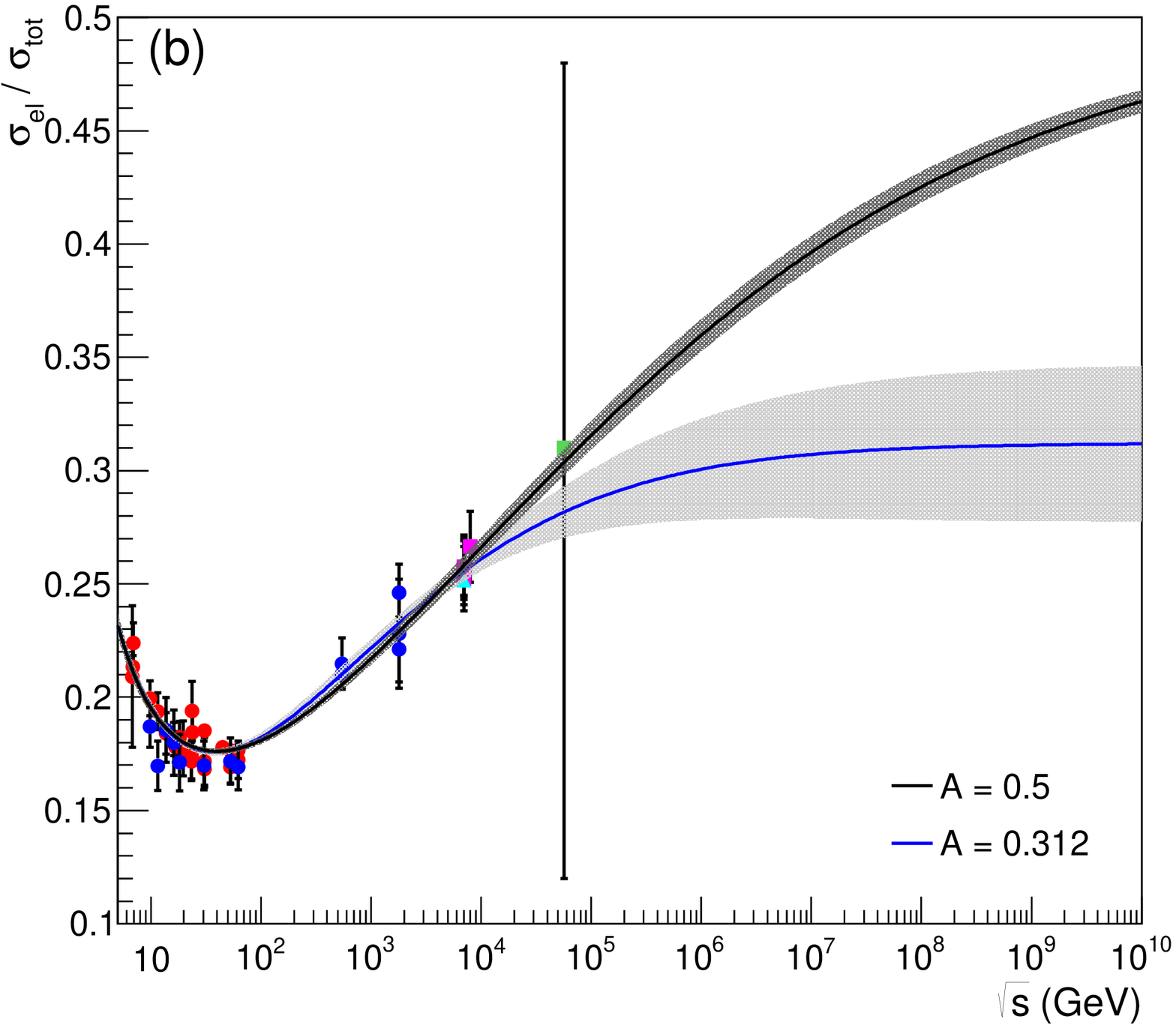,width=8cm,height=7cm}
\caption{Selected fit results for $X(s)$ with variant $LL$ (Eq. (\ref{lll})), $A$ free (gray),
$A$ fixed (black) and sigmoid: (a) logistic; (b) tanh.}
\label{f6}
\end{figure}

Summarizing, in what follows we concentrate on the results
obtained with variant $LL$, sigmoid logistic and tanh and two scenarios,
for short referred to as gray ($A$ free) and black ($A$ = 0.5).
For comparison, in Fig. \ref{f6}
we display all these results for $X(s)$ together with the corresponding uncertainty
regions and with the energy region extended up to $10^{10}$ GeV. 
Numerical predictions at some energies of interest for $X(s)$ (and other ratios to
be discussed in what follows) are shown in the third columns of Table \ref{t12} (logistic, $A$ = 0.293
and $A$ = 0.5) and Table \ref{t13} (tanh, $A$ = 0.312 and $A$ = 0.5).

\begin{table}[!ht]
\caption{Predictions for different ratios at the LHC energy region
and beyond, with the \textit{logistic}, variant $LL$, $A$ = 0.293 (gray)
and $A$ = 0.5 (black).} 
\begin{center}
\begin{tabular}{llllll} 
\hline
$A$ &$\sqrt{s}$ (TeV)&$\sigma_{el}/\sigma_{tot}$&$\sigma_{inel}/\sigma_{tot}$&$\sigma_{diff}/\sigma_{tot}$&
$\sigma_{tot}/B$ \\
\hline
      & 2.76   &  0.2408(37) & 0.7592(37) & 0.2592(37) & 12.11(19) \\
      &   8    &  0.2575(49) & 0.7425(49) & 0.2425(49) & 12.94(25) \\
 0.293&   13   &  0.2637(65) & 0.7363(65) & 0.2363(65) & 13.26(33) \\  
      &   57   &  0.278(13)  & 0.722(13)  & 0.222(13)  & 13.95(64) \\
      &   95   &  0.281(15)  & 0.719(15)  & 0.219(15)  & 14.11(74) \\
\hline
      & 2.76 & 0.2366(32) & 0.7634(32) & 0.2634(32) & 11.89(16) \\
      &   8    & 0.2625(44) & 0.7375(44) & 0.2375(44) & 13.20(22) \\
 0.5  &   13   & 0.2750(49) & 0.7250(49) & 0.2250(49) & 13.82(25) \\
      &   57   & 0.3141(64) & 0.6859(64) & 0.1859(64) & 15.79(32) \\
      &   95   & 0.3276(69) & 0.6724(69) & 0.1724(69) & 16.46(34) \\  
\hline
\end{tabular}
\end{center}
\label{t12}
\end{table}

\begin{table}[!ht]
\caption{Predictions for different ratios at the LHC energy region
and beyond, with the \textit{tanh}, variant $LL$, $A$ = 0.312 (gray) and $A$ = 0.5 (black).} 
\begin{center}
\begin{tabular}{llllll}
\hline
$A$ &$\sqrt{s}$ (TeV)&$\sigma_{el}/\sigma_{tot}$&$\sigma_{inel}/\sigma_{tot}$&$\sigma_{diff}/\sigma_{tot}$&
$\sigma_{tot}/B$ \\
\hline
     &2.76 & 0.2404(36) & 0.7596(36) & 0.2596(36) & 12.08(18) \\
     &   8    & 0.2577(46) & 0.7423(46) & 0.2423(46) & 12.95(23) \\
0.312&   13   & 0.2647(58) & 0.7353(58) & 0.2353(58) & 13.30(29) \\
     &   57   & 0.282(11)  & 0.718(11)  & 0.218(11)  & 14.16(57) \\
     &   95   & 0.286(13)  & 0.714(13)  & 0.214(13)  & 14.39(67) \\
\hline
     &2.76 & 0.2381(33) & 0.7619(33) & 0.2619(33) & 11.97(17) \\
     &   8    & 0.2612(42) & 0.7388(42) & 0.2388(42) & 13.13(21) \\
0.5  &   13   & 0.2719(46) & 0.7281(46) & 0.2281(46) & 13.66(23) \\
     &   57   & 0.3038(57) & 0.6962(57) & 0.1962(57) & 15.27(28) \\
     &   95   & 0.3145(60) & 0.6855(60) & 0.1855(60) & 15.81(30) \\
\hline
\end{tabular}
\end{center}
\label{t13}
\end{table}

\begin{figure}[h!]
\centering
\epsfig{file=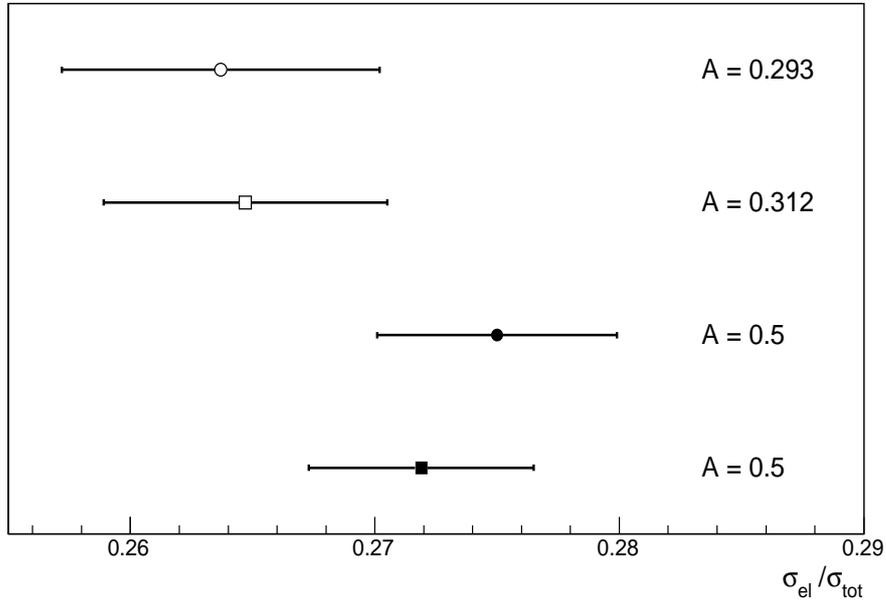,width=12cm,height=8cm}
\caption{Preditions for the ratio $X$ at 13 TeV, with variant $LL$, gray ($A$ free) and black
($A$ = 0.5) scenarios and sigmoids logistic (circles)
and tanh (squares).}
\label{f7}
\end{figure}

These predictions deserve some comments.
From Fig. \ref{f6}, in the case of $A$ free and taking into account the uncertainty regions,
the extrapolations are consistent with our asymptotic value 0.303 $\pm$ 0.055 above
$\sqrt{s} \approx$ 10$^3$ TeV. In other words the results suggest that the asymptotic region
might already be reached around 10$^3$ TeV. That, however, is not the case with the black disk
for which typical asymptopia are predicted far beyond 10$^{10}$ TeV.
In what concerns Run 2, let us focus on the predictions for $X(s)$
at 13 TeV. The four results from Tables \ref{t12} and \ref{t13} are schematically
summarized in Figure \ref{f7}. Although the expected measurements at this energy
might not strictly discriminate between the two scenarios it may possible to have an indicative
in favor of one of them. Even if that were not the case, these new experimental information will 
certainly be
crucial for a better determination of the curvature beyond the inflection point.  

\section{Extension to Other Quantities}
\label{s6}

With the selected results, let us present extensions
and predictions to some other quantities of interest,
that did not take part in the data reductions.

\subsection{Inelastic Channel: Ratios and Diffractive Dissociation}

\begin{figure}[h!]
\centering
\epsfig{file=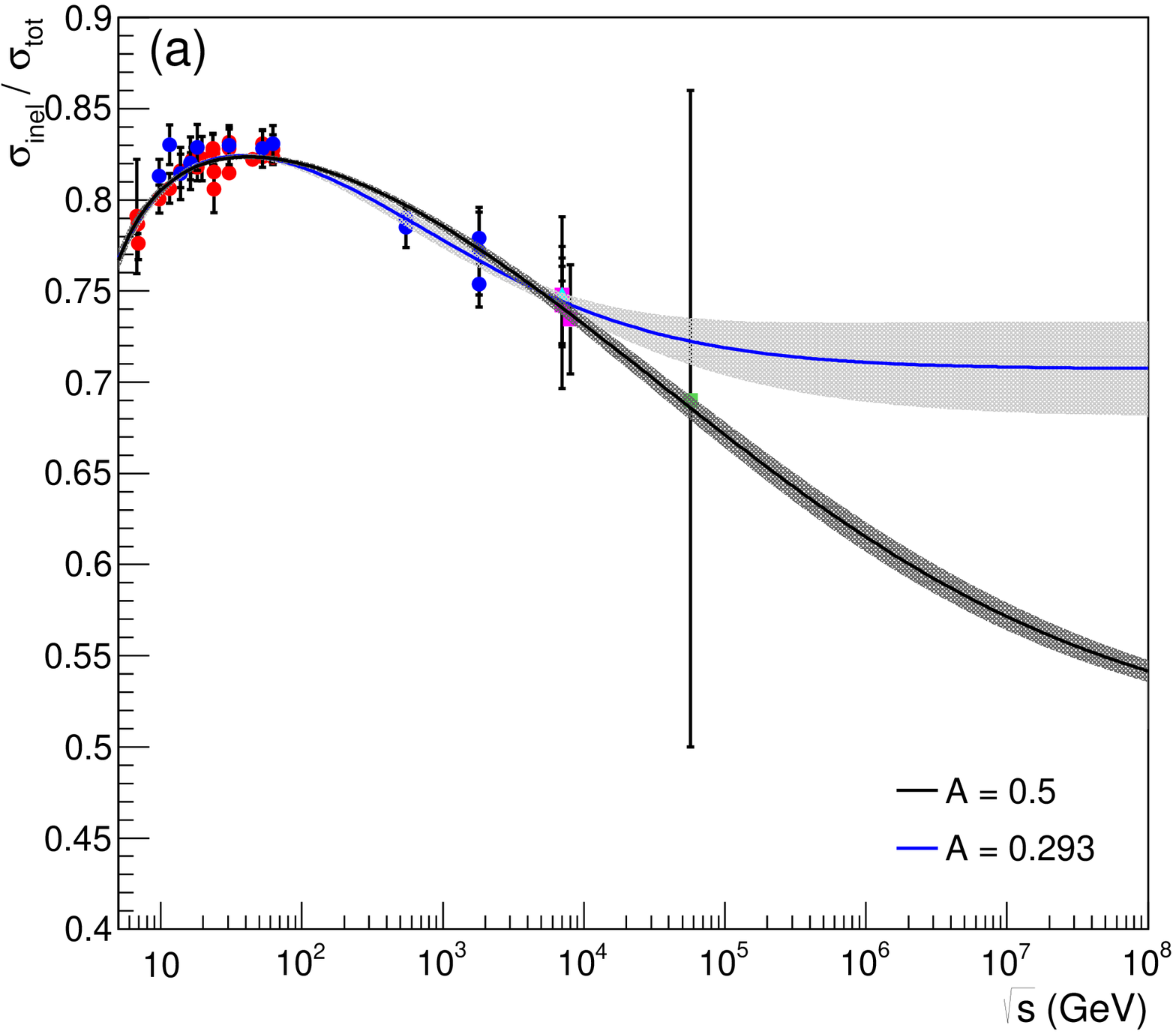,width=8cm,height=7cm}
\epsfig{file=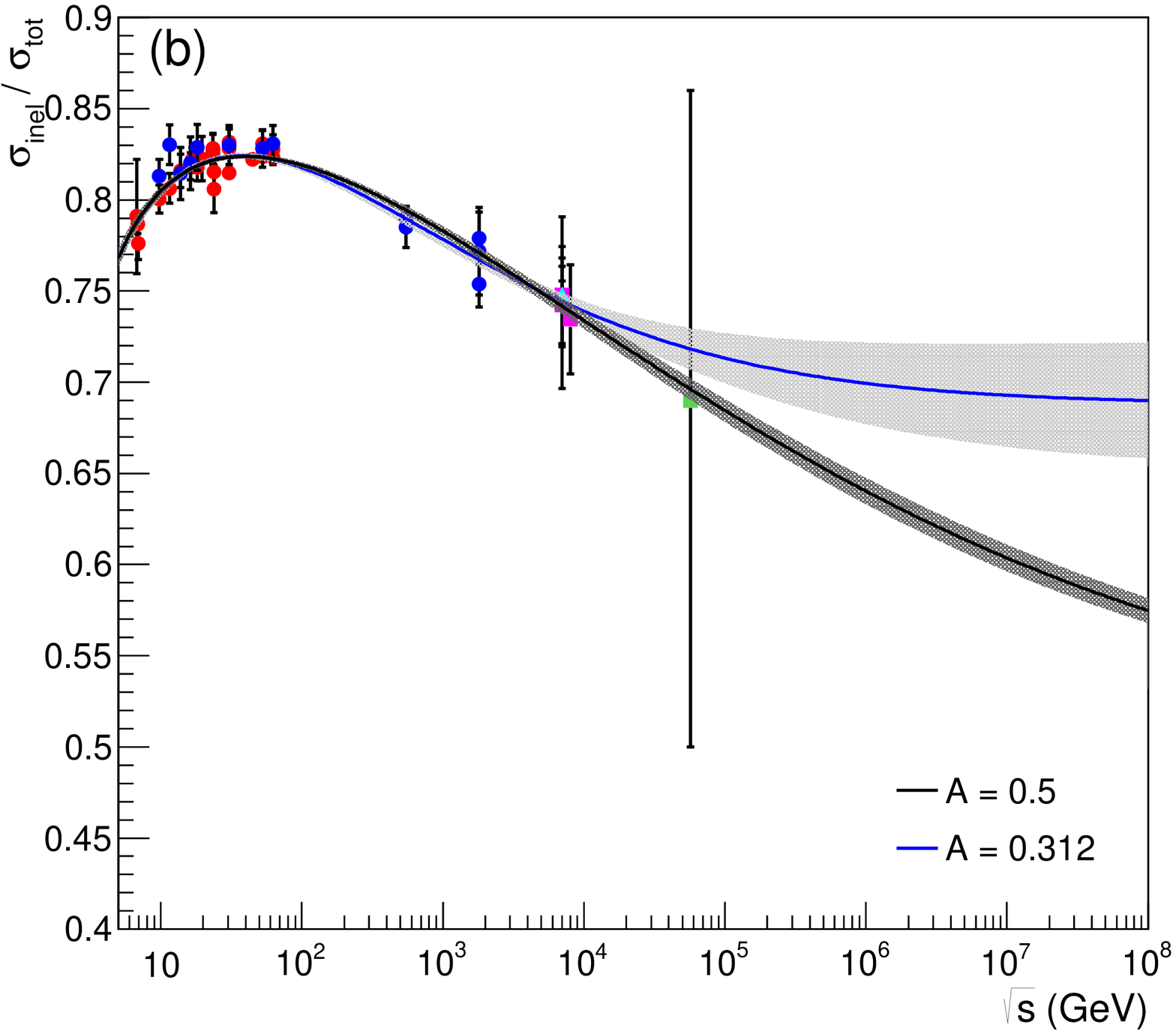,width=8cm,height=7cm}
\caption{ Predictions, via unitarity, for the ratio between the inelastic and total cross 
sections, with variant $LL$ (Eq. (\ref{lll})), $A$ free (gray),
$A$ fixed (black) and sigmoids: 
(a) logistic; 
(b) tanh.}
\label{f8}
\end{figure}

The ratio between the inelastic and total cross-section is directly
obtained via unitarity, namely 1 - $X(s)$. The results (analogous
to those for $X(s)$ in Fig. \ref{f6}) are displayed in Fig. \ref{f8},
together with the uncertainty regions and the experimental data available.
As expected, all cases present consistent descriptions of the
experimental data.
Numerical predictions at the energies of interest for the ratio
$\sigma_{inel}/\sigma_{tot}$
are shown in the fourth columns of Table \ref{t12} (logistic, $A$ = 0.293
and $A$ = 0.5) and Table \ref{t13} (tanh, $A$ = 0.312 and $A$ = 0.5).

In what concerns the asymptotic limit,
\begin{eqnarray}
\lim_{s \rightarrow \infty} \frac{\sigma_{inel}}{\sigma_{tot}} = 1 - A,
\nonumber
\end{eqnarray}
contrasting with the black disk (1/2), our global ($g$) and restricted ($r$) estimations,
Eqs. (\ref{g}) and (\ref{r}), predict
\begin{eqnarray}
1 - A_g = 0.70 \pm 0.12
\qquad
\mathrm{and}
\qquad
1 - A_r = 0.697 \pm 0.055.
\nonumber
\end{eqnarray}

Beyond elastic scattering, the soft inelastic diffractive  processes
(single and double dissociation) play a fundamental role in the
investigation of the hadronic interactions. An important formal result
on the diffraction dissociation cross-section concerns the Pumplin 
upper bound \cite{pump1,pump2}:
\begin{equation}
 \frac{\sigma_{el}}{\sigma_{tot}} + \frac{\sigma_{diff}}{\sigma_{tot}} \leq \frac{1}{2},
\nonumber
 \label{eq:pumplin}
\end{equation}
where $\sigma_{diff}$ stands for the soft inelastic diffractive cross section
(the sum of the single and double dissociation cross sections).
In this context, the black disk limit (1/2) may be associated with a combination
of the soft processes, namely elastic and diffractive, giving room, therefore
to a semi-transparent scenario.

In this respect, Lipari and Lusignoli have recentely discussed the experimental data
presently available on these processes, calling the attention to the possibility
that the Pumplin bound may already be reached at the LHC energy region \cite{lipari2}. 
The argument is based on
a combination of the measurements by the TOTEM and ALICE Collaborations at 7 TeV,
which indicates
\begin{eqnarray}
\frac{\sigma_{diff}}{\sigma_{tot}} \approx 0.24_{-0.06}^{+0.05},
\qquad
\frac{\sigma_{el} + \sigma_{diff}}{\sigma_{tot}} = 0.495_{-0.06}^{+0.05},
\qquad
\frac{\sigma_{diff}}{\sigma_{el}} = 0.952_{-0.24}^{+0.20},
\nonumber
\end{eqnarray}
suggesting therefore that the Pumplin bound is close to saturation.

In case of saturation, namely the equality
in the above equation, it is possible to estimate the ratio $\sigma_{diff}/\sigma_{tot}$
at the LHC energies and beyond. The numerical predictions
for this ratio
are shown in the fifth columns of Table \ref{t12} (logistic, $A$ = 0.293
and $A$ = 0.5) and Table \ref{t13} (tanh, $A$ = 0.312 and $A$ = 0.5).
Obviously, the asymptotic value of this ratio is zero in the case of a black disk scenario strictly
associated with the elastic channel.

\begin{figure}[h!]
\centering
\epsfig{file=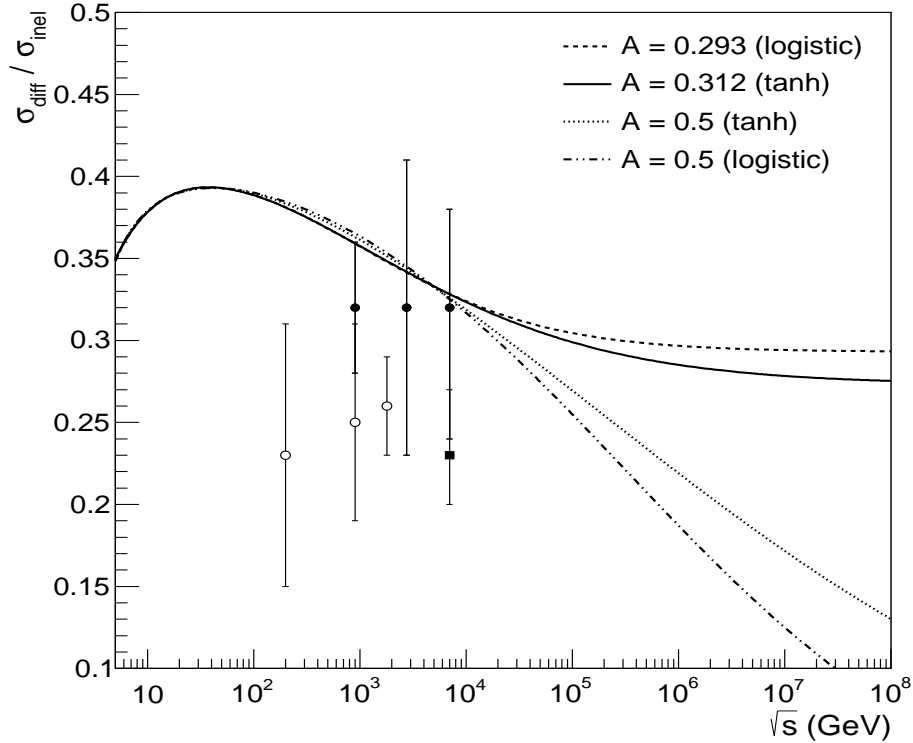,width=12cm,height=10cm}
\caption{Upper bounds for the ratio between the dissociative (single plus double) and inelastic cross sections 
from the selected results for
the gray and black scenarios (logistic and tanh). Experimental information from $\bar{p}p$ scattering
at 0.2, 0.9 and 1.8 TeV (white marks) \cite{ua51,ua52,cdf1,cdf2,cdf3} and $pp$ scattering 
at 0.9, 2.76 and 7 TeV (black marks) \cite{alice,cms1,cms2}.}
\label{f9}
\end{figure}

Moreover, using the Pumplin bound and Unitarity, we can also infer an upper bound for 
ratio $\sigma_{diff}/\sigma_{inel}$, namely
\begin{equation}
R(s) \equiv \frac{\sigma_{diff}}{\sigma_{inel}} \leq \frac{1/2 - X(s)}{1-X(s)}.
\nonumber
\end{equation}
The curves corresponding to this bound in all cases treated in this Section are shown in Fig. 
\ref{f9}, together with experimental data.
Once more, contrasting with the asymptotic null limit in a black disk scenario, our
global and reduced estimations ($A_g$ and $A_r$),  lead to the predictions:

\begin{equation}
R_g = 0.29 \pm 0.12
\qquad
\mathrm{and}
\qquad
R_r = 0.283 \pm 0.055
\qquad
\mathrm{for}
\qquad
s \rightarrow \infty.
\nonumber
\end{equation}

\subsection{Ratio Y Associated with Total Cross-Section and Elastic Slope}

In cosmic-ray studies, the determination of the $pp$ total cross-section from the proton-air
production cross-section is based on the Glauber formalism \cite{engel1,engel2,engel3}.
In this context, the nucleon-nucleon impact parameter amplitude
(profile function) constitutes an important ingredient for the connection between hadron-hadron and
hadron-nucleus scattering.
This function is typically parametrized by 
\begin{eqnarray}
a_j(s,\vec{b}_j) = \frac{[1 + \rho(s)]}{4\pi}\, \frac{\sigma_{tot}(s)}{B(s)}\, 
e^{-\vec{b}_j^{2}/[2B(s)]},
\end{eqnarray}
where $\rho$, $\sigma_{tot}$ and $B$
demand inputs from models to complete the connection. However, models have been tested
only in the accelerator energy region and in general are characterized by  different
physical pictures and different predictions at higher energies. As a consequence, the 
extrapolations result in large theoretical
uncertainties, as clearly illustrated by Ulrich \textit{et al}. \cite{engel1}. 
From the above equation, any extrapolation is strongly dependent on the ratio
$\sigma_{tot}/B$,
namely the unknown correlation between $\sigma_{tot}$ and $B$ in terms of energy.

One way to overcome this difficulty is to estimate the ratio
\begin{equation}
Y(s) = \frac{\sigma_{tot}}{B}(s)
\end{equation}
through its approximate relation with the ratio $X(s)$, treated in \ref{saa},
\begin{eqnarray}
Y(s) \approx 16 \pi X(s).
\end{eqnarray}
That has been the motivation
of the analysis presented in \cite{fm12,fm13}:
``an almost model-independent parametrization for the above ratio
may reduce the uncertainty band in the extrapolations from accelerator to 
cosmic-ray-energy regions." 

\begin{figure}[h!]
\centering
\epsfig{file=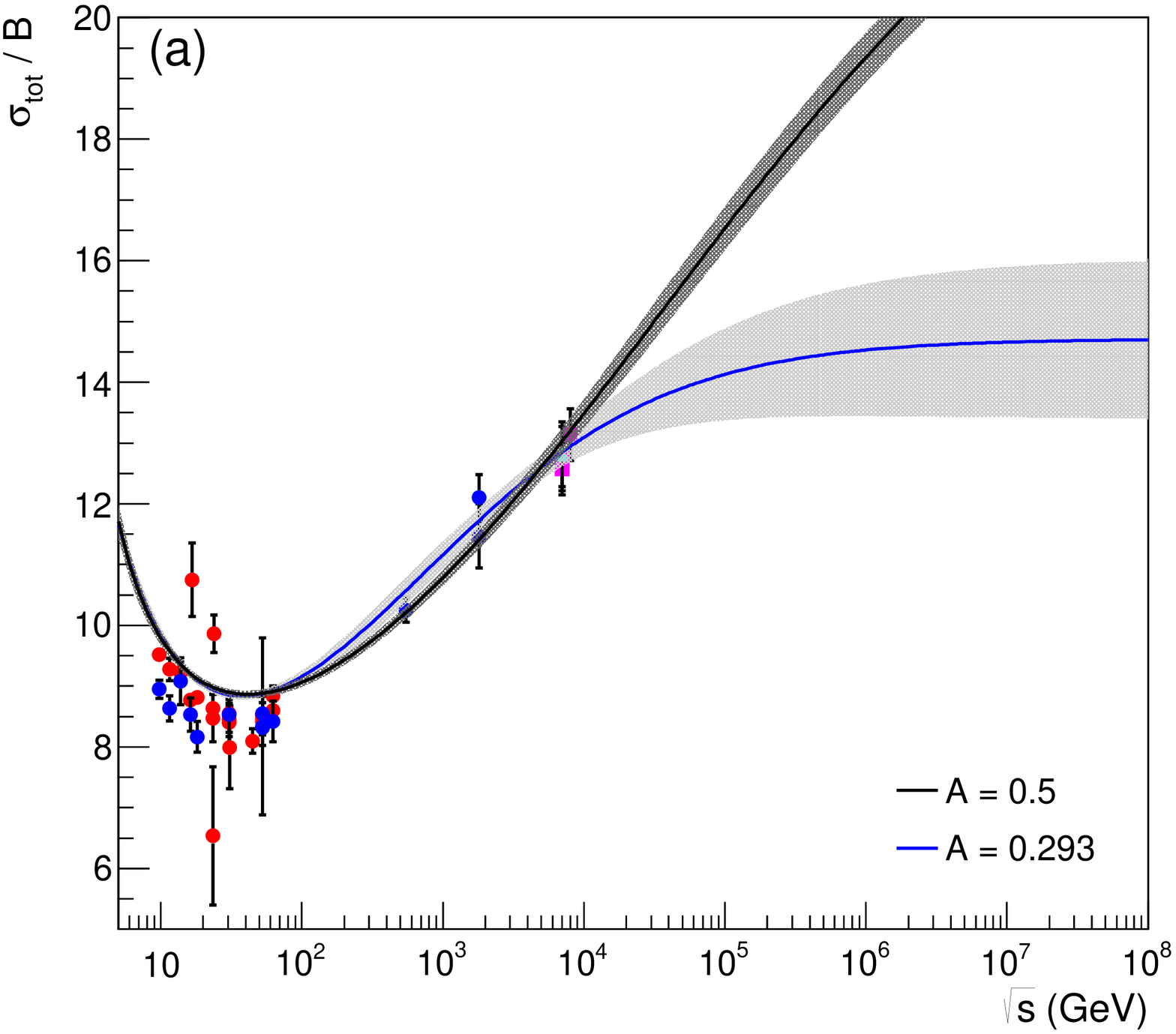,width=8cm,height=7cm}
\epsfig{file=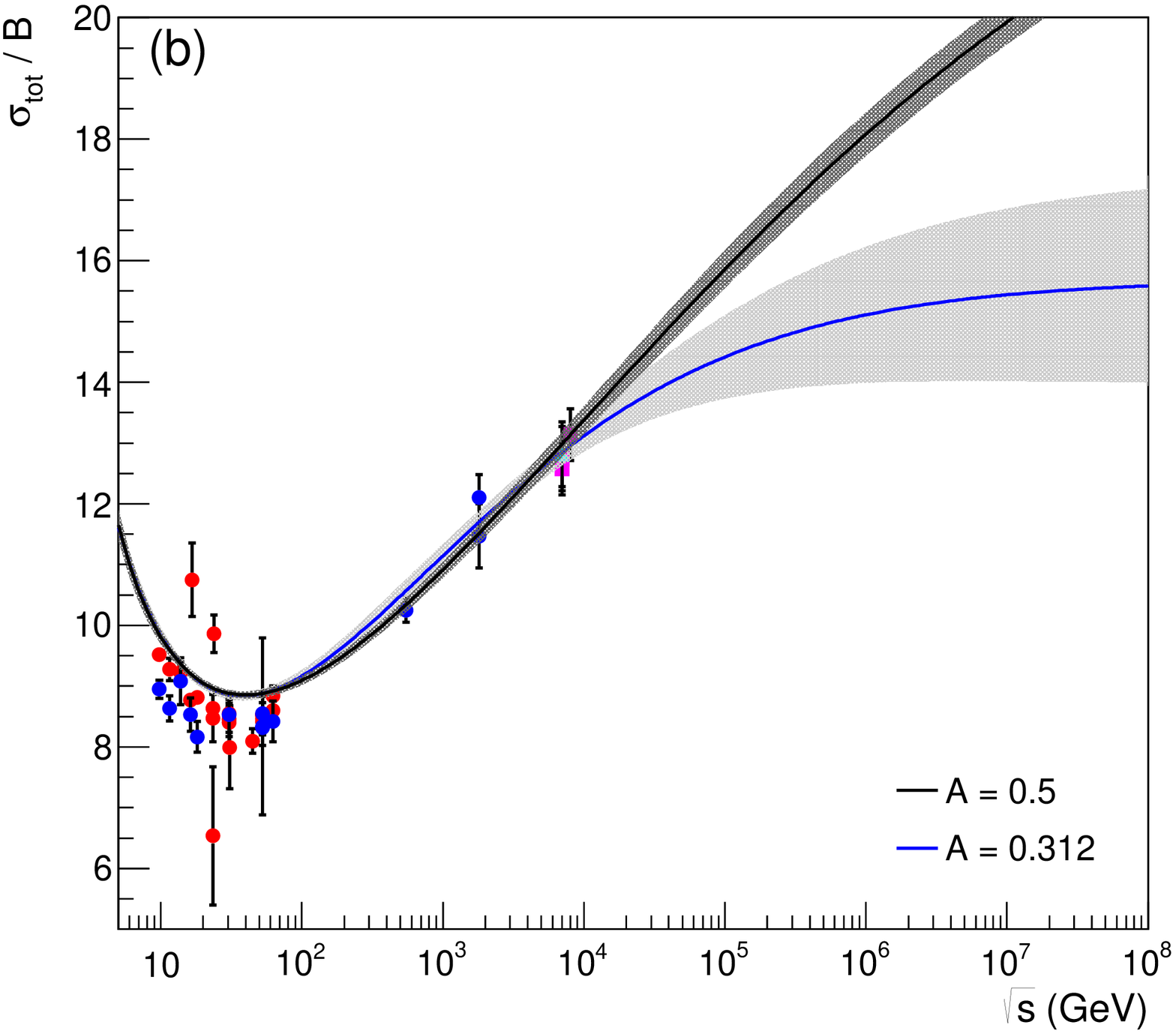,width=8cm,height=7cm}
\caption{ Predictions for the ratio between the total cross 
sections and the elastic slope, with variant $LL$, (Eq. (\ref{lll})), $A$ free (gray),
$A$ fixed (black) and sigmoid: 
(a) logistic; 
(b) tanh.}
\label{f10}
\end{figure}

The behavior of $Y(s)$ extracted in this way and in all cases treated in this Section
are shown in Fig. \ref{f10}; the corresponding numerical predictions
at the energies of interest
are displayed in the sixth columns of Table \ref{t12} (logistic, $A$ = 0.293
and $A$ = 0.5) and Table \ref{t13} (tanh, $A$ = 0.312 and $A$ = 0.5).

\section{Further Comments}
\label{s7}

In this section we present some additional comments on two topics:
possible physical interpretations associated with our empirical
parameterizations (Section 7.1) and aspects related to a semi-transparent (or gray)
asymptotic scenario (Section 7.2).

\subsection{On Possible Physical Interpretations}

As we have shown, the sigmoid and elementary functions selected for
parameterizing $X(s)$, represent suitable and efficient choices on 
both statistical and phenomenological grounds. Moreover, a striking feature
is associated with the small number of free parameter involved: from 3 to 5, depending 
on the procedure. 
Despite the strictly empirical character of our analysis and strategies,
let us attempt to connect the results provided by our sigmoid ansatz to some reasonable 
physical concepts.
The goal is not a direct interpretation of the results, but to discuss some ideas
that could contribute with further investigation on the subject.
This discussion is not related to a specific asymptotic scenario but
focus only on our choices for the sigmoid and elementary functions. 

We discuss two aspects: (1) the possible connection of $X(s)$ with 
the contribution of effective partonic interactions through the opacity
concept; (2) the relation of this process with a saturation effect
associated to a change of curvature in $X(s)$.

Beyond technology and exact sciences, sigmoid functions \cite{sig1,sig2} have 
applicability in a great variety 
of areas, as biological, humanities and social sciences, including neural networks,
language change, diffusion of an innovation and many others
\cite{sig3,sig4,sig5}. Application in analytic unitarization schemes (eikonal/U-matrix) has been
considered by Cudell, Predazzi, Selyugin \cite{cps1,cps2} and a recent application 
related to polarized gluon density is discussed by Bourrely \cite{bourrely}.

This class of functions is generally associated with the Pearl-Verhulst logistic
processes, in which the growth of a population is bounded and proportional to its
size, as well as, to the difference between the size and its bound. The process may be
represented by the logistic differential equation \cite{sig3,sig4}
\begin{equation}
\frac{dN}{dt} = r N \left[1 - \frac{N}{M}\right],
\end{equation}
where $N = N(t)$ is the population at a time $t$, $M$ is its maximum value 
(the carrying capacity) and $r > 0$ is the intrinsic rate of growth
(the growth rate per capita). If $r=r(t)$ and/or $M=M(t)$ the equation is
with variable coefficients \cite{sig5}.

In our case, the logistic ansatz for $S(f)$, Eq. (\ref{l}), is a trivial solution of the
differential equation
\begin{equation}
\frac{dS}{df} = S \left[1 - S\right].
\end{equation}
In terms of the ratio $X$ and variable $v$, Eqs. (\ref{xafs}) and (\ref{v}) respectively,
this differential equation reads
\begin{equation}
\frac{dX}{dv} = \frac{df}{dv} X \left[1 - \frac{X}{A}\right],
\end{equation}
corresponding, therefore, to a logistic equation with variable
coefficient $df/dv$.

How to interpret this result in the context of elastic hadronic
scattering at high energies? One conjecture
is to look for possible connections with an effective number of
partonic interactions taking part in the collision processes,
as the energy increases.

With that in mind, we recall that the wide class of QCD inspired models,
presently in the literature, had its bases in previous concepts related
to semi-hard QCD or mini jet models \cite{chl,prec1,prec2,lmv,ps,dp3,capella,deus}.
The main ingredient concerned the separation of soft and semi-hard contributions
in the scattering process; the former treated on phenomenological grounds
and the latter determined under perturbative QCD, parton model and probabilistic arguments.

Specifically, as recalled in \ref{saa}, in terms
of the opacity function ($\Omega$), the probability that an inelastic event takes place at $b$ and $s$
is given by
\begin{equation}
G_{inel}(s,b) = 1 - e^{-2\Omega}.
\end{equation}
Denoting the probability that there are \textit{no} soft (semi-hard) inelastic interaction by
$\bar{P}_S$ ($\bar{P}_{SH}$), we can associate
\begin{equation}
G_{inel}(s,b) = 1 - \bar{P}_S \bar{P}_{SH} = 1 - e^{-2\Omega_S} e^{-2\Omega_{SH}},
\label{gi} 
\end{equation}
so that, the total opacity reads
\begin{equation}
\Omega(s,b) = \Omega_S(s,b) + \Omega_{SH}(s,b).
\end{equation}

Let us now focus on the semi-hard opacity, which is constructed under probabilistic arguments
and QCD parton model as follows.
Let $n(s,b)$ be the \textit{number of parton-parton collisions at $s$ and $b$, which
is associated with the probability of semi-hard inelastic scattering}. Since
$\bar{P}_{SH} = e^{- 2\Omega_{SH}(s,b)}$ is the probability that there are $no$ 
semi-hard inelastic interactions, mean-free path arguments lead to
the conclusion that the probability that hadrons do not undergo SH
scattering can be expressed by
\begin{eqnarray}
 \bar{P}_{SH} = e^{- n(s,b)} \nonumber
\end{eqnarray}
and therefore, from Eq. \ref{gi}, the semi-hard opacity reads
\begin{equation}
\Omega_{SH}(s,b) = \frac{1}{2} n(s,b).
\end{equation}

In the theoretical context, $n(s,b)$ is expressed in terms of parton-parton cross section
and hadronic matter distribution (related to form factors in the $q^2$-space). In general,
the formalism involves a rather complex structure with different choices for
the corresponding $\Omega_S$, $\Omega_{SH}$, constituent cross-sections and
form factors (see, for example, \cite{fagfrasc} for a very recent version). 

However, for our purposes, the main idea is that an effective number
of partonic interactions is clearly connected with the opacity and
the profile functions and therefore, as recalled in \ref{saa},  with the ratio $X(s)$
(at least in what concerns the central opacity in gray disk and Gaussian profiles).
In this context, the sigmoid behavior characterized by a change of curvature,
suggests a change in the rate of effective partonic interactions taking
part in the hadronic collisions. A fast rise at low energies is followed
by a saturation effect starting at the inflexion point, which represents
a change in the dynamics of the interaction.

The energy $\sqrt{s}$ of the inflection point
in terms of the asymptotic ratio $A$ and in all cases investigated
is displayed in Tab. \ref{t11}.
We see that, despite the different values of $A$, the position of
the inflection point lies in a rather restrict interval,
namely $\sqrt{s} \approx$ 80 - 100 GeV. 

Concerning this region, we recall that the UA1 Collaboration have
reported the measurement of low transverse energy clusters (mini jets) 
in $\bar{p}p$ collisions at the CERN Collider and $\sqrt{s}$: 200 - 900 GeV
\cite{ua1}. Extrapolation of the observed mini jet cross-section
to lower energy (Fig. 10 in \cite{ua1}) suggest that the 
region 80 - 100 GeV is consistent with the beginning of the mini jet production.
Now, the rise of the mini jet cross-section has been associated with
the observed faster rise of the inelastic and, consequently the total cross-section
(\cite{ua1} and references there in). Therefore, once $X = \sigma_{el}/\sigma_{tot}$, 
it seems reasonable to associate this behavior with a change of curvature in $X$ 
and the beginning of a saturation effect. Our results are consistent with
this conjecture.

Although suggestive, we stress that the above arguments are certainly limited 
as effective physical interpretations of our analysis and results.
However, we hope they may be useful as a possible context for further 
investigation.

\subsection{On a Semi-Transparent Asymptotic Scenario}

From the discussion in Section \ref{s5}, our analysis favors
the global result $A_g = 0.3 \pm 0.12$; a scenario below the black disk. 
Let us discuss some results and aspects related to the possibility of a 
semi-transparent scenario.
 
Up to our knowledge, in the mid-seventies, Fia\l{}kowski and Miettinen had already
suggested a semi-transparent scenario in the context of a multi-channel approach \cite{miet}; 
based on the Pumplin bound, this scenario has been also conjectured by 
Sukhatme and Henyey \cite{suk}.
More recently, the advent of the LHC brought new expectations 
concerning the historical search for asymptopia, but interpretation
of the data seems not easy in the phenomenological context.
In respect to scenarios below the black disk in the pre and LHC era,
we recall the facts that follows.

\vspace{0.3cm}

\noindent
1. Lipari and Lusignoli \cite{lipari1} and Achilli et al. \cite{achilli} have discussed
the observed overestimation of $\sigma_{el}$ (or underestimation of $\sigma_{inel}$)
in the context of one channel eikonal models. That led Grau et al. \cite{grau} to re-interpret
the Pumplin upper bound as an effective asymptotic limit,
\begin{eqnarray}
\frac{\sigma_{el}}{\sigma_{tot}} + \frac{\sigma_{diff}}{\sigma_{tot}}
\rightarrow \frac{1}{2}
\qquad
\mathrm{as}
\qquad
s \rightarrow \infty.
\nonumber
\end{eqnarray}
Based on the behavior of the experimental data on $X(s)$ and a combination 
of the results at 57 TeV by the Auger Collaboration together with the
predictions by Block and Halzen, Grau et al.
have conjectured a rational limit $A$ = 1/3 as a possible asymptotic value
\cite{grau}; therefore, consistent with our preferred results.

\vspace{0.3cm}

\noindent
2. In the recent QCD motivated model by Kohara, Ferreira and Kodama 
\cite{kfk1,kfk2,kfk3,kfk4}
the scattering amplitude is
 constructed
under both perturbative and non-perturbative QCD arguments (related to extensions
of the Stochastic Vacuum Model). The model leads to consistent descriptions
of the experimental data on $\bar{p}p$ and $pp$ elastic scattering (forward quantities
and differential cross sections) above 20 GeV, 
including the LHC energies and with
extensions to the cosmic-ray energy region. The predictions for the asymptotic
ratio $X$ lie below 1/2 and are close to 1/3; therefore in agreement with our
selected results.

\vspace{0.3cm}

\noindent
3. As discussed before, combination of the parameterizations by the COMPETE ($\sigma_{tot}$) and TOTEM 
($\sigma_{el}$) Collaborations indicates the asymptotic value $A$ = 0.436.
Moreover, combination of the ATLAS parametrization ($\sigma_{el}$) \cite{atlas} with
that by COMPETE reads $A$ = 0.456. Both, therefore, below the black disk.

\vspace{0.3cm}

\noindent
4. At last, we stress that through a completely different approach, several individual and simultaneous fits to 
$\sigma_{tot}$ and $\rho$ data, extended to fit the $\sigma_{el}$
data \cite{fms13,ms13a,ms13b}, have led to asymptotic ratios in plenty agreement
with our inferred global limit $A_g = 0.30 \pm 0.12$.

\vspace{0.3cm}

All the above facts corroborate the results and conclusions 
presented here, indicating the semi-transparent limit as a possible
asymptotic scenario for the hadronic interactions.

At last, it may be instructive to recall a crucial aspect related
to our subject. Besides empirical analysis on the ratio $X$, another
important model-independent way to look for empirical information on the 
central opacity is through the inverse scattering problem,
namely empirical fits to the differential cross section data and
inversion of the Fourier transform connecting the scattering
amplitude and the profile function (\ref{saa}). In this context,
the Amaldi and Schubert analysis on the $pp$ data from the CERN-ISR
(23.5 $\leq \sqrt{s} \leq$ 62.5 GeV) constitute a classical example
of extraction of the inelastic overlap function \cite{as}.

Still restricted to $pp$ scattering at the ISR, but including data from the
Fermilab at 27.4 GeV,  several characteristics of the
profile, inelastic and opacity function (in both impact parameter
and momentum transfer space) have been extracted and discussed
by Menon and collaborators in references \cite{cm,cmm,sma,am,fms,fm}. 
The approach is characterized by analytical model-independent parameterizations
for the scattering amplitude and analytical propagation of the uncertainties
from the fit parameters to all the extracted quantities. An important observation from
these analyses is the necessity of adequate experimental information in 
the region of large momentum transfer in order to obtain reliable 
results on the central region (small impact parameter).
This crucial fact was already pointed out by R. Lombard in the 
first Blois Meeting on Elastic  
and Diffractive Scattering (1985):
".. extrapolating the measured differential cross section can be made
in an infinite number of manners. Some extrapolated curves may look
unphysical, but they cannot be excluded on mathematical grounds." \cite{lombard}.

The experimental data from the ISR on $pp$ collisions 
cover the region in momentum transfer up to 9.8 GeV$^2$ and with the
inclusion of the data from the Fermilab, this region can be extended
up to 14 GeV$^2$ \cite{am}, allowing therefore consistent empirical analysis
(see, for example, Fig. 7 in \cite{cmm} or Fig. 1 in \cite{sma} and the detailed
discussions in \cite{am}).
That, however, is not the case with the $\bar{p}p$ data (ISR, S$\bar{p}p$S
Collider or Tevatron) and, unfortunately, with the $pp$ data from the LHC
neither:
the TOTEM measurements cover only the region up to 2.5 or 3.0 GeV$^2$. 
As a consequence, propagation of the uncertainties from the extrapolated
fits led to rather large uncertainty regions at small values of the
impact parameter.
It is interesting to notice that this effect has been clearly pointed out
in the theoretical papers by  Khoze, Martin and Ryskin (see, for example, Fig. 2 in
\cite{kmr} and related comments).

It is also important to notice that even discrete Fourier transforms,
through adequate bins intervals in momentum transfer, can not guarantee
the consistency of the extracted profile near the central region.
Recall that one of the main controversy concerning differential cross
sections at large momentum is related to the possibility of oscillation
or smooth decrease and/or presence or not of a second dip. 
Obviously, all these important features (connected with the central region),
are completely lost in empirical fits restricted to small or medium
values of the momentum transfer (a fact that is implicit in the 
aforementioned statement by Lombard). 

On the bases of the above discussion, we understand that only experimental data
in the \textit{deep-elastic scattering region} (above $\approx$ 4 GeV$^2$ \cite{deep})
can provide the necessary information for reliable empirical
extraction of the profile function (and opacity) at and near
the central region. In that sense, it would be very important if the
experiments at the LHC could extend the region of momentum transfer.
In this respect Kawasaki, Maehara and Yonezawa,
stated in 2003 \cite{kmy} (also quoted in
\cite{cmm}): ``Such experiments will give much more valuable information for the
diffraction interaction rather than to go to higher energies".

\section{Summary, Conclusions and Final Remarks}
\label{s8}

We have presented an empirical analysis on the energy dependence 
of the ratio $X = \sigma_{el}/\sigma_{tot}$, with focus on its
asymptotic limit. The main ingredient concerned four
analytical parameterizations constructed through composition of
sigmoid and elementary functions of the energy. For each sigmoid, $S$
(logistic or hyperbolic tangent), two elementary functions, $f$, were considered,
given by a linear function plus a power law ($PL$) or a logarithmic law ($LL$) of the standard variable
$\ln s/s_0$, with $s_0 = 4m_p^2$. By expressing $X(s) = A S(f(s))$, with $A$ the
asymptotic limit, two fit procedures have been considered, either constrained
($A$ fixed, imposing the asymptotic limit) or unconstrained ($A$ as a free parameter, selecting
the asymptotic limit). Based on empirical and formal arguments, we have investigated
5 limits from $A$ = 0.3 up to $A$ = 1 (maximum unitarity), including the black disk case, $A$ = 0.5. 
Altogether, we have developed 40 data reductions, associated with 5 asymptotic
assumptions, 2 sigmoid functions composed with 2 elementary functions and 2 procedures (unconstrained
and constrained). The dataset comprised all the experimental data available on the
ratio $X$ above 5 GeV (and up to 8 TeV).
This detailed empirical analysis led to two main results:

\begin{description}

\item{R1.} 
Taking into account the statistical information and uncertainties,
all \textit{constrained} fits ($A$ fixed) presented consistence with the experimental data
analyzed, even in the extrema cases, $A$ = 0.3 or $A$ = 1.0.
In other words, all the scenarios investigated are equally probable and our
analysis does not allow to select a particular scenario (from the constrained fits). 

\item{R2.} Independently of the sigmoid or elementary function considered, all the 
unconstrained fits ($A$ free) converged to an unique solution within the uncertainties,
which is also consistent with the experimental data analyzed. Based on the four
results, a global asymptotic value can be inferred: $A_g = 0.30 \pm 0.12$;
moreover, a restricted result to variant $LL$ indicates $A_r = 0.303 \pm 0.055$.

\end{description}

These two results led us to two main conclusions:

\begin{description}

\item{C1.} 
Although consistent with the experimental data analyzed, the black disk limit
does not represent an unique or exclusive solution.

\item{C2.}  
Our analysis favors a limit below the black disk, namely a gray or semi-transparent
scenario, with $A$ given above.

\end{description}

Based on these two conclusions and given the ubiquity of the black disk scenario
in eikonal models,
we have presented along the text the results and several predictions
considering both the black and semi-transparent cases with the logistic and hyperbolic tangent
and restricted to our variant $LL$ 
(since it presents the smaller number of free parameters).

An important characteristic distinguishing the two scenarios
concerns the approach to the asymptotic region. In the semi-transparent scenario
asymptopia may already be reached around 10$^3$ TeV and in the case
of the black disk only far beyond 10$^{10}$ TeV.

In what concerns our empirical parameterizations, it seems important
to stress their efficiency in describing the experimental data analyzed,
independently of the fixed physical value of $A$ and 
despite the noticeable small number of free parameters. For example,
the constrained black disk fit with the variant $LL$ demands only three
fit parameters and led to consistent description of all data.
We stress, once more, the contrast with 10 or more parameters typical
of individual fits to $\sigma_{tot}$ and $\sigma_{el}$ data.
All that indicate the
good quality of our analytical choices for $S(f)$ and $f(s)$, on empirical
and phenomenological grounds. 

Given the empirical efficiency of these \textit{analytical representations} 
for $X(s)$, we have attempted to look for possible physical connections
with the underline theory/phenomenology of  the soft strong interactions.
Presently, as a first step, we can only devise some suggestive ideas
relating a ``population growth", represented by the logistic
differential equation, with the number of effective parton interactions
taking part in the collision. This number
increases at low energies, tending to a saturation as
the energy increases above the 80 - 100 GeV region. Despite
suggestive, this conjecture is rather limited on physical grounds 
and must be investigated in more detail.

It seems difficult to consider as accidental the semi-transparent
scenario indicated in a variety of independent analyses as those
presented here, as well as in our previous works \cite{fms15a,fms15b,fms13,ms13a,ms13b}
and other indications referred to in the text.

It is worth mentioning that an asymptotic scenario distinct of the black disk
has been recently discussed by I. Dremin, in the context of unitarity and a 
Gaussian profile  approach \cite{d1,d2,d3}. In this analysis
the opacity is maximum at a finite value of the impact parameter and the
central opacity is determined by the parameter $Z \equiv 4 \pi B/\sigma_{tot}$.
The asymptotic limit $Z \rightarrow 0.5$ corresponds to complete transparency,
leading the author to propose a black torus (or a black ring) as a possible
asymptotic scenario. Since, from Appendix A, $Z(s) \approx 1/4X(s)$, with our
preferred restricted asymptotic result $A_r$ (19), we predict 
\begin{eqnarray}
Z \approx 0.83
\qquad
\mathrm{as}
\qquad
s \rightarrow \infty,
\nonumber
\end{eqnarray}
corresponding, therefore, to a central semi-transparent core and not complete transparency.
This asymptotic scenario, represented by a black ring surrounding a gray
disk, has been also predicted by Desgrolard, Jenkovszky and Struminsky,
in a different context \cite{djs1,djs2}.

At last, we stress that the new experimental data from Run 2
might not be decisive as a direct numerical selection of the scenarios discussed 
here (in terms of our predictions). However, this experimental information 
on $X(s)$  will be crucial for an improved  determination of the curvature
above the inflection point and consequently providing better accesses
to the asymptotic region in model-independent analyses.

\vspace{0.3cm}

\section*{Acknowledgments}

Research supported by FAPESP, Contract 2013/27060-3 (P.V.R.G.S.).

\appendix

\section{Basic Formulas and Results}
\label{saa}

In this appendix we collect some formulas and results characteristic of the elastic
hadron scattering, which are referred to along the text. We treat two subjects:
the eikonal and impact parameter representations (Section A.1) and the
approximate relation connecting the ratio between the elastic and total cross section
with the ratio between the total cross section and the elastic slope (Section A.2).
The text and notation are based on references \cite{pred,dremin,bc}.

\subsection{Impact Parameter and Eikonal Representations}

\subsubsection{Physical Quantities}

In elastic hadron scattering, the amplitude $F$ is usually expressed
as a function of the Mandesltam variables $s$ and $t = - q^2$. 
In terms of this amplitude the physical quantities of interest here,
with the corresponding normalization, are the differential cross section
\begin{eqnarray}
\frac{d\sigma}{dq^2} = \pi |F(s, q)|^2,
\label{ea1}
\end{eqnarray}
the elastic (integrated) cross section,
\begin{eqnarray}
\sigma_{el}(s) = \int_0^{\infty} \frac{d\sigma}{dq^2} dq^2,
\label{ea2}
\end{eqnarray}
the total cross section (optical theorem)
\begin{eqnarray}
\sigma_{tot}(s) = 4 \pi Im F(s, q=0),
\label{ea3}
\end{eqnarray}
the inelastic cross section (via unitarity),
\begin{eqnarray}
\sigma_{inel}(s) = \sigma_{tot}(s) - \sigma_{el}(s)
\label{ea4}
\end{eqnarray}
and the $\rho$ parameter,
\begin{eqnarray}
\rho(s) = \frac{Re\,F(s, q=0)}{Im\,F(s,q=0)}.
\label{ea5}
\end{eqnarray}

From the above formulas, the optical point reads
\begin{eqnarray}
\left.\frac{d\sigma}{dq^2} \right|_{q^2=0} = \frac{\sigma_{tot}^2[1 + \rho^2]}{16 \pi}.
\label{ea6}
\end{eqnarray}

\subsubsection{Impact Parameter Representation}

The representation of the scattering amplitude in the impact parameter space
is named \textit{profile function} (denoted $\Gamma$) and in case of azimuthal symmetry, they
are related by
\begin{eqnarray}
F(s, q) = i \int_0^{\infty} b\, db\, J_0(qb)\, \Gamma(s,b).
\label{ea7}
\end{eqnarray}

The unitarity principle in the impact parameter space is usually expressed in terms
of the total, elastic and inelastic overlap functions,
$G_{tot}(s, b) = G_{el}(s, b) + G_{inel}(s, b)$, which, in terms of the profile function reads
\begin{eqnarray}
2 Re \Gamma(s,b) = |\Gamma(s,b)|^2 + G_{inel}(s,b).
\label{ea8}
\end{eqnarray}
In this representation the total, elastic and inelastic cross sections
are given, respectively  by 
\begin{eqnarray}
\sigma_{tot}(s) = 4 \pi \int_0^{\infty} b\, db\, Re \Gamma(s,b),
\qquad
\sigma_{el}(s) = 2 \pi \int_0^{\infty} b\, db\, |\Gamma(s,b)|^2, 
\label{ea9}
\end{eqnarray}
\begin{eqnarray}
\sigma_{inel} = 2 \pi \int_0^{\infty} b\, db\, G_{inel}(s,b).
\label{ea10}
\end{eqnarray}

\subsubsection{Eikonal Representation and the Opacity Function}

In the eikonal representation, the profile function is expressed by
\begin{eqnarray}
\Gamma(s,b) = 1 - e^{i\,\chi(s,b)},
\label{ea11}
\end{eqnarray}
where $\chi(s,b)$ is the complex valued \textit{eikonal function}. In this representation,
from the unitarity relation Eq. (\ref{ea8}),
\begin{eqnarray}
G_{inel}(s,b) = 1 - e^{- 2 Im\,\chi(s,b)}.
\label{ea12}
\end{eqnarray}
Since unitarity implies $Im\,\chi (s,b) \geq 0$, we have
$G_{in}(s, b) \leq $1, so that from Eq. (\ref{ea10}), $G_{in}(s,b)$ can be interpreted as the probability
of an inelastic event to take place at given $b$ and $s$. 
From this result, the imaginary part of the eikonal is associated with
the absorption in the scattering process and for that reason (and the
optical analogy), it is
named \textit{opacity function}, which we shall denote
\begin{eqnarray}
\Omega(s,b) \equiv \mathrm{Im} \chi(s,b).
\label{ea13}
\end{eqnarray}

Let us neglect the real part of the scattering amplitude, so that $\Gamma(s,b)$
is a real valued function. In this case,
\begin{eqnarray}
\Gamma(s,b) = 1 - e^{- \Omega(s,b)},
\end{eqnarray}
and by expanding the exponential term, in first order,
\begin{eqnarray}
\Gamma(s,b) \approx \Omega(s,b).
\end{eqnarray}
Therefore, the profile is also connected with the hadronic opacity.
In fact, in the optical analogy (Fraunhofer diffraction),
the function $\eta = e^{- \Omega(s,b)} = 1 - \Gamma(s,b)$ represents the modification
of the incident wave caused by a diffracting object. In this context, without
the object, $\eta=1$ and $\Gamma=0$ (no diffraction) and in case of a
completely opaque object, $\eta=0$ and $\Gamma=1$ (no transmission).

\subsubsection{Gray Disk, Black Disk and Gaussian Profiles}

From the above discussion, a real valued profile function for a gray disk of
radius $R(s)$ and central opacity $\Gamma_0(s) = \Gamma(s, b=0)$, is
represented by

$$
\Gamma(s,b) =
\left\{\begin{array}{ll}
 \Gamma_0(s), & 0 \leq b \leq R(s)\\
 0, &  b > R(s)
\end{array}\right.
$$
From Eqs. (\ref{ea9}),  $\sigma_{tot} = 2 \pi \Gamma_0(s) R^2(s)$
and $\sigma_{el} =  \pi \Gamma_0^2(s) R^2(s)$, so that the ratio $X$
is given by

\begin{eqnarray}
X_{GD} = \frac{\Gamma_0(s)}{2},
\end{eqnarray}
and the case of a black disk ($\Gamma_0(s) = 1$) reads
\begin{eqnarray}
X_{BD} = \frac{1}{2}.
\end{eqnarray}

A Gaussian profile, with central opacity $\Gamma_0(s)$,
\begin{eqnarray}
\Gamma(s, b) = \Gamma_0(s) e^{-b^2/R^2},
\nonumber
\end{eqnarray}
is important because, through Eqs. (\ref{ea1}), (\ref{ea2}) and (\ref{ea6}) and (\ref{ea7}), it can
describe the sharp forward peak in the differential cross section data,
leading to the experimental determination of the integrated elastic cross section
\cite{totem1,atlas} (see the next Subsection A.2). In this case, one obtains 
$\sigma_{tot} = 2 \pi \Gamma_0(s) R^2(s)$
and $\sigma_{el} =  \pi \Gamma_0^2(s) R^2(s) / 2$, so that
\begin{eqnarray}
X_{G}(s) = \frac{\Gamma_0(s)}{4}.
\end{eqnarray}

Therefore, these  examples provide an interpretation of the
ratio $X(s)$ as the energy dependence of the central opacity (or blackness)  
associated with the  interacting hadrons.

\subsection{The $X$ and $Y$ Ratios}

The experimental data on the differential cross section is characterized
by the dominance of a sharp forward peak, which can be well described by
\begin{eqnarray}
\frac{d\sigma}{dq^2} = \left.\frac{d\sigma}{dq^2} \right|_{q^2=0}\, e^{-Bq^2},
\end{eqnarray}
with $B$ the (constant) forward slope. Substituting in the optical point,
Eq. (\ref{ea6}), and then integrating Eq. (\ref{ea2}),
the elastic cross section reads
\begin{eqnarray}
\sigma_{el}(s) = \frac{[1 + \rho^2]}{B(s)}\, \frac{\sigma_{tot}^2(s)}{16 \pi}.
\nonumber
\end{eqnarray}
Since, from the experimental data $\rho(s) \lesssim \mathrm{0.14}$,
to take $1 + \rho^2 \approx \mathrm{1}$ is a reasonable approximation, leading to
\begin{eqnarray}
\frac{\sigma_{tot}(s)}{B(s)} \approx 16 \pi \frac{\sigma_{el}(s)}{\sigma_{tot}(s)},
\end{eqnarray}
or, with our notation,
\begin{eqnarray}
Y(s) \approx 16 \pi X(s).
\end{eqnarray}

\section{Short Review of Previous Results}
\label{sab}

In this Appendix, using the notation defined in Section \ref{s3},
we review some 
previous results we have obtained in fits with the tanh, special
cases of the variant $PL$ and energy scales at 1 and 25 GeV$^2$.

In the 2012 analysis by Fagundes and Menon \cite{fm12,fm13}, the dataset were restricted to $pp$
scattering above 10 GeV and included only the first TOTEM datum at 7 TeV. In order to infer
uncertainty regions in the extrapolation to higher energies, two extreme asymptotic
limits have been tested by either fixing $A$ = 1/2 (black-disk limit) or $A$ = 1 (maximum
unitarity). The experimental data have been well described through variant $PL$ with fixed $\delta = 2$, 
fixed $s_0 = 1$ GeV$^2$,  and only three fit parameters: $\alpha$, $\beta$ and $\gamma$
(denoted $\gamma_1$, $\gamma_2$ and $\gamma_3$ in \cite{fm12}).
Through the approximate relation $Y \approx 16\pi X$ (Appendix A.2), it was possible
to extend the extrapolation of the uncertainty regions to the ratio $\sigma_{tot}/B_{el}$, 
which, in the context of the Glauber model \cite{fm12}, plays an important role in the determination
of the proton-proton total cross-section from proton-air production cross-section in cosmic-ray experiments.

In our subsequent analysis \cite{fms15a,fms15b}, the energy cutoff has been extended down to 5 GeV
and the dataset included all the TOTEM measurements at 7 and 8 TeV \cite{fms15a}, as well as the recent
ATLAS datum at 7 TeV \cite{fms15b}. Preliminary fits to only $pp$ data with variant $PL$, energy
scale as the energy cutoff ($s_0 = 25$ GeV$^2$) and different $A$ values led 
to an almost unique solution indicating the parameter $\delta$ consistent with
0.5, within the uncertainties. We than fixed $\delta = 0.5$ and developed new fits
including now the $\bar{p}p$ data, considering either constrained and unconstrained
cases and different values for the $A$ parameter. All data reductions presented consistent
descriptions of the experimental data analyzed and in the case of the unconstrained fit
we have obtained an unique solution with asymptotic limit below the black disk, namely 
$A = 0.332 \pm 0.049$ \cite{fms15b}.


\end{document}